\documentclass[12pt]{article}
\usepackage{latexsym, amsfonts, amsmath}

\usepackage[active]{srcltx}
\usepackage{graphicx}

\newtheorem{thm}{Theorem}[section]

\newtheorem{lem}[thm]{Lemma}

\newtheorem{exam}[thm]{Example}
\newtheorem{rem}[thm]{Remark}

\def\infint{\int_{-\infty}^\infty}

\def\convd{\stackrel{\cal D}{\rightarrow}}

\def\I{{\rm I}}
\def\ex{{\rm E\,}}
\def\Cov{{\rm Cov\, }}
\def\Var{{\rm Var\, }}

\def\R{\mathop{\mathbb R}\nolimits}
\def\var{\mathop{\rm Var}\nolimits}

\begin{document}

\bibliographystyle{plain}

\title {Bivariate Uniform Deconvolution}

\author {   Martina Bene\v{s}ov\'{a}, Bert van Es*  and  Peter Tegelaar\\[0.3cm]
{\normalsize * Korteweg-de Vries Institute  for Mathematics}\\
{\normalsize University of Amsterdam}\\
{\normalsize Science Park 904,
 1018 TV Amsterdam,}\\
{\normalsize P.O. Box  94248, 1090 GE Amsterdam}\\
{\normalsize The Netherlands}}


\maketitle

\begin{abstract}
We construct  a density estimator
in the   bivariate uniform deconvolution model. For this model we derive four inversion formulas
to express the bivariate density that we want to estimate in terms of the bivariate density of the observations.
By substituting a kernel density estimator of the density of the observations we then get
four different estimators. Next we construct an asymptotically optimal convex combination of these
four estimators.
Expansions for the bias, variance, as well as asymptotic normality, are derived.
Some simulated examples are presented.
\\[.5cm]

\end{abstract}

\section{Introduction}

Before focusing on bivariate deconvolution let us first consider    univariate deconvolution .
Let $X_1,\ldots, X_n$ be i.i.d. observations, where $ X_i=Y_i+Z_i $
and $Y_i$ and $Z_i$ are independent.  Assume that the unobservable
$Y_i$ have distribution function $F$ and density $f$.   Also assume that
the unobservable random variables $Z_i$ have a known density $k$.
If the $Z_i$ are uniformly distributed then we
have a {\em uniform deconvolution problem}. Note that the
density $g$ of $X_i$ is equal to the convolution of $f$ and $k$, so $g=k*f$ where
$*$ denotes convolution.  So we have
\begin{equation}\label{convgen}
g(x)=\int_{-\infty}^\infty k(x-u)f(u)du.
\end{equation}
The deconvolution problem is the problem of estimating $f$ or $F$
from the observations $X_i$.

Several generally applicable methods have been proposed for this deconvolution model.
The   standard {\em Fourier type kernel density estimator}  for deconvolution
problems is based on the Fourier transform, see for instance Wand
and Jones (1995).
Let $w$ denote a {\em kernel function} and $h>0$ a {\em
bandwidth}.
The estimator $f_{nh}(x)$ of the density $f$ at the point $x$ is defined as
\begin{equation}
f_{nh}(x)=\frac{1}{2\pi} \int_{-\infty}^\infty
e^{-itx} \frac{\phi_w(ht)\phi_{emp}(t)}{ \phi_k(t)}\,dt=
{1\over nh}\sum_{j=1}^n v_h\Big({{x-X_j}\over h}\Big),
\end{equation}
with
$$
v_h(u)={1\over 2\pi}\infint {{\phi_w(s)}\over\phi_k(s/h)}\
e^{-isu}ds,\quad\mbox{and}\quad
\phi_{emp}(t) = {1\over n}\sum_{j=1}^n e^{itX_j},
$$
the empirical characteristic function, and $\phi_w$ and $\phi_k$
denote the characteristic functions of $w$ and $k$ respectively. An
important condition for these estimators to be properly defined is
that the characteristic function $\phi_k$ of the density $k$ has no
zeroes, which renders it useless for uniform deconvolution. In fact,
Hu and Ridder (2004) argue that in economic applications this
assumption is not reasonable since many distributions with a bounded
support have characteristic functions with zeros on the real line.
They propose an approximation of the Fourier transform estimator in
such cases. For other modifications of the Fourier inversion method
in this problem see Hall and Meister (2007),Feuerverger, Kim and Sun
(2008), Meister (2008) and Delaigle and Meister (2011).

In some univariate deconvolution problems one can apply {\em nonparametric maximum likelihood}.
In the uniform deconvolution problem for instance the error $Z$ is
Uniform$[0,1)$ distributed. So in this particular deconvolution
problem we assume to have i.i.d. observations from the density
\begin{equation}\label{convun}
g(x)=\int_{-\infty}^\infty I_{[0,1)}(x-u)f(u)du =\int_{x-1}^x f(u)du = F(x)-F(x-1).
\end{equation}
Groeneboom and Jongbloed (2003) consider density estimation in this
problem. They propose a kernel density estimator based on the
nonparametric maximum likelihood estimator (NPMLE) of the
distribution function $F$ and derive its asymptotic properties. For
estimators of the distribution function in uniform deconvolution,
related to the NPMLE,  we refer to Groeneboom and Wellner (1992),
Van Es and Van Zuijlen (1996) and Donauer, Groeneboom and Jongbloed
(2009).

A selected group of deconvolution problems allows explicit {\em inversion formulas} of
(\ref{convgen}) expressing the density of interest $f$ in terms
 of the density $g$ of the data. In these cases we can
estimate $f$ by substituting for instance a direct kernel density
estimate of $g$ in the inversion formula. In Van Es and Kok (1998)
this strategy has been pursued for deconvolution problems where
$k$ equals the exponential density, the Laplace density, and their
repeated convolutions.

If we apply inversion to the uniform problem then it turns out we get two obvious inversion formulas.
Of course these inversions agree on the set of
densities of the form (\ref{convun}), but they are different outside of this set.
Plugging in a kernel estimator of the density $g$  of the observations, which is typically {\em not} of this form, then yields two estimators
of $f$. These can then in some sense be optimally combined in a convex combination. This approach is developed in Van Es (2011).  Here we will follow this approach in the bivariate uniform deconvolution setting.

\bigskip

Let us now consider  {\em bivariate deconvolution}.
The bivariate convolution formula $\mathbf{X}_i = \mathbf{Y}_i + \mathbf{Z}_i$, where $\mathbf{X}_i, \mathbf{Y}_i$ and $\mathbf{Z}_i$ stand for two dimensional random vectors, can be written in vector notation as
\begin{equation}
    \begin{pmatrix}
    X_{i1}\\
    X_{i2}
    \end{pmatrix} =
    \begin{pmatrix}
    Y_{i1}\\
    Y_{i2}
    \end{pmatrix} +
    \begin{pmatrix}
    Z_{i1}\\
    Z_{i2}
    \end{pmatrix}.
\end{equation}
The estimation principles described above can in principle all be attempted in the bivariate problem as well. See for instance Youndj\'e and Wells (2008) for recent results on multivariate Fourier type kernel deconvolution.   Approaches based on nonparametric maximum likelihood and inversion hardly exist to our knowledge.

In the bivariate uniform deconvolution setting the random vector $\mathbf{Z}_i$ has a Uniform$([0,1)\times[0,1))$ distribution, i.e. it is uniformly distributed on the unit square.
Here we can also express the bivariate density $g$ of the observations in terms of the bivariate distribution function $F$, with density $f$, of the random vector $\mathbf{Y}$. We have
\begin{align}
    g(x_1,x_2) &= \int_{-\infty}^\infty \int_{-\infty}^\infty I_{[0,1)}(x_1-u_1)I_{[0,1)}(x_2-u_2)f(u_1,u_2)du_1 du_2\nonumber\\
    &= \int_{x_2-1}^{x_2} \int_{x_1-1}^{x_1} f(u_1,u_2)du_1 du_2\nonumber\\
    &= F(x_1,x_2) - F(x_1,x_2-1) - F(x_1-1,x_2) + F(x_1-1,x_2-1).\label{convunbi}
\end{align}
This is the bivariate analogue of formula (\ref{convun}). Note  that, again,  the Fourier inversion approach can not be used because of the zeros in the characteristic function of
the bivariate uniform distribution.

\bigskip

Apart from being of theoretical interest, bivariate unform
deconvolution is also of interest because of its relation to what
one might call {\em quadrant censoring} or {\em bivariate current
status data}, i.e. a bivariate version of univariate Type I interval
censoring. This censoring problem can be described as follows. For
convenience we restrict ourselves to the unit square. Consider $n$
i.i.d random points ${\mathbf T_i}, i=1,\ldots,n$, with ${\mathbf
T_i}=(T_{i1},T_{i2})$, in the unit square. Furthermore we have $n$
i.i.d unobservable random points ${\mathbf X_i}, i=1,\ldots,n$, with
${\mathbf X_i}=(X_{i1},X_{i2})$, also in the unit square. For each $i$ we
observe whether ${\mathbf X_i}$ is in quadrant $1,2,3$ or $4$
relative to the known point ${\mathbf T_i}$. Let us quantify these
observations by the discrete random variable $\Delta_i$. So we have
\begin{equation}
\Delta_i= \left\{
\begin{array}{ll}
1 &,\ \mbox{if}\ X_{i1}\geq T_{i1}\ \mbox{and}\ X_{i2}\geq T_{i2},\\
2 &,\ \mbox{if}\ X_{i1}<T_{i1}\ \mbox{and}\ X_{i2}\geq T_{i2},\\
3 &,\ \mbox{if}\ X_{i1}<T_{i1}\ \mbox{and}\ X_{i2}< T_{i2},\\
4 &,\ \mbox{if}\ X_{i1}\geq T_{i1}\ \mbox{and}\ X_{i2}<T_{i2}.\\
\end{array}
\right.
\end{equation}
This problem is related to uniform deconvolution by a tranformation of the data.
Assume that the unobserved ${\mathbf X_i}$ have a bivariate
density $f$. The statistical problem is to estimate this density from the
observations $({\mathbf T_1},\Delta_i),\ldots, ({\mathbf
T_n},\Delta_n)$.

Consider the following transformation of the points ${\mathbf T_i}$,
\begin{equation}{\mathbf V_i}=(V_{i1},V_{i2})=
\left\{
\begin{array}{ll}
(T_{i1}+1,T_{i2}+1) &,\ \mbox{if}\ \Delta_i=1,\\
(T_{i1},T_{i2}+1) &,\ \mbox{if}\ \Delta_i=2,\\
(T_{i1},T_{i2}) &,\ \mbox{if}\ \Delta_i=3,\\
(T_{i1}+1,T_{i2}) &,\ \mbox{if}\ \Delta_i=4.\\
\end{array}
\right.
\end{equation}
It can be shown that if the density $f$ is concentrated on the unit
square and if the observation points $\mathbf T_i$ are uniformly
distributed on the unit square then the density of the random points
$\mathbf V_i$ is identical to (\ref{convunbi}). This shows that a
method for bivariate uniform deconvolution of the type developed
here can also be used in quadrant censoring.

\bigskip

The main aim of this paper is to develop the inversion approach of Van Es (2011)  for   bivariate uniform deconvolution.
In Chapter \ref{inversion} we derive four inversion formulas for (\ref{convunbi}). This yields the same number of possible estimators if we plug in a density estimator of the density $g$ of the observations. In Chapter \ref{estimation} we combine these estimators in a convex combination which is asymptotically optimal in some sense. The weights of this combination turn out to depend on the unknown distribution $F$. A general theorem for an estimator with estimated weights is given in Chapter \ref{estweights}.
We also present specific estimators of these weights. Simulated examples  are presented in Chapter \ref{simulation}.
Chapter \ref{proofs} contains  the proofs.

\section{Inversion  formulas}\label{inversion}

Recall that the density of the $\mathbf{Z}_i$ is equal to $k(z_1,z_2)=I_{[0,1)\times[0,1)}(z_1,z_2)=I_{[0,1)}(z_1)I_{[0,1)}(z_2)$. This yields formula (\ref{convunbi}) which expresses $g(x_1,x_2)$ in terms of $F(x_1,x_2)$.
Lemma \ref{Inversion} below demonstrates  that the converse is also feasible.

First note that for
\begin{align*}
&F^{--}(y_1,y_2):=\Pr(Y_1 \leq y_1,Y_2 \leq y_2),\\
&F^{-+}(y_1,y_2):=\Pr(Y_1 \leq y_1,Y_2 > y_2),\\
&F^{+-}(y_1,y_2):=\Pr(Y_1 > y_1,Y_2 \leq y_2),\\
&F^{++}(y_1,y_2):=\Pr(Y_1 > y_1,Y_2 > y_2).
\end{align*}
\medskip
the following equalities hold
\medskip
\begin{align}
    &F^{--}(x_1,x_2) = F(x_1,x_2),\label{eq1}\\
    &F^{-+}(x_1,x_2) = F_{Y_1}(x_1) -F(x_1,x_2) ,\label{eq2}\\
    &F^{+-}(x_1,x_2) =  F_{Y_2}(x_2) -F(x_1,x_2),\label{eq3}\\
    &F^{++}(x_1,x_2) = F(x_1,x_2) - F_{Y_1}(x_1) - F_{Y_2}(x_2) + 1 \label{eq4}.
\end{align}
If we know $F(x_1,x_2)$ and if this function is continuously differentiable over $x_1$ and $x_2$, then we know $f(x_1,x_2)$, because $f(x_1,x_2) = \frac{\partial^2}{\partial x_1\partial x_2} F(x_1,x_2)$. In fact,   combined with the formulas above, and (\ref{convunbi}), this gives us four different inversion formulas to obtain $f$ and $F$ from $g$, as is stated in the following Lemma.

\begin{lem}\label{Inversion}
We have
\begin{align}
&F^{--}(x_1,x_2) = \sum_{i=0}^\infty \sum_{j=0}^\infty g(x_1-i,x_2-j),\label{--}\\
&F^{-+}(x_1,x_2) = \sum_{i=0}^\infty \sum_{j=1}^\infty g(x_1-i,x_2+j),\label{-+}\\
&F^{+-}(x_1,x_2) = \sum_{i=1}^\infty \sum_{j=0}^\infty g(x_1+i,x_2-j),\label{+-}\\
&F^{++}(x_1,x_2) = \sum_{i=1}^\infty \sum_{j=1}^\infty g(x_1+i,x_2+j).\label{++}
\end{align}
Assume that $\lim_{x_1\to\pm\infty}f(x_1,x_2)=0$ and
$\lim_{x_2\to\pm\infty}f(x_1,x_2)=0$. Furthermore, assume that
$g(x_1,x_2)$ is twice mixed continuously differentiable over $x_1$
and $x_2$. Then there are four inversion formulas to recover $f$
from $g$. We have
\begin{align}
&f(x_1,x_2) = \sum_{i=0}^\infty \sum_{j=0}^\infty \frac{\partial^2}{\partial x_1\partial x_2} g(x_1-i,x_2-j),\\
&f(x_1,x_2) = -\sum_{i=0}^\infty \sum_{j=1}^\infty \frac{\partial^2}{\partial x_1\partial x_2} g(x_1-i,x_2+j),\\
&f(x_1,x_2) = -\sum_{i=1}^\infty \sum_{j=0}^\infty \frac{\partial^2}{\partial x_1\partial x_2} g(x_1+i,x_2-j),\\
&f(x_1,x_2) = \sum_{i=1}^\infty \sum_{j=1}^\infty \frac{\partial^2}{\partial x_1\partial x_2} g(x_1+i,x_2+j).
\end{align}
\end{lem}

\bigskip

To get some more insight in these inversion formulas  note that (\ref{convunbi}) can be interpreted as a probability for
$\mathbf{Y}$ (under $F$). We have
$$
g(x_1,x_2)= P_F(\mathbf{Y}\in(x_1-1,x_1]\times(x_2-1,x_2]).
$$
So $g(x_1,x_2)$ is equal to the probability that $\mathbf{Y}$ belongs to a specific square $(x_1-1,x_1]\times(x_2-1,x_2]$.
Adding up over suitable squares we then get the probability that $\mathbf{Y}$ belongs to a specific quadrant with a given vertex. For a formal proof see Chapter \ref{proofs}.

\begin{figure}[h]
$$\includegraphics[width=7cm]{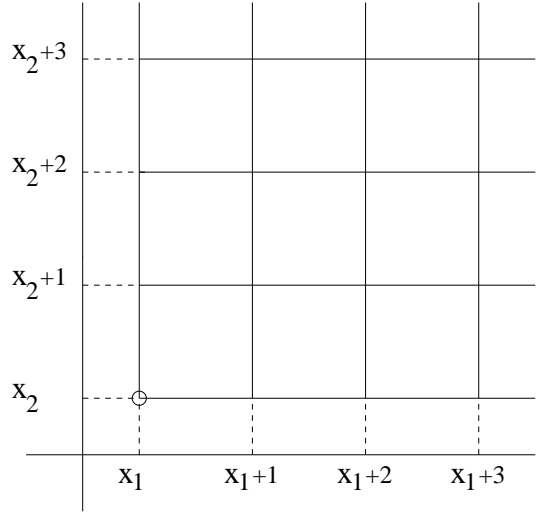}$$
\caption[]{ $F^{++}(x_1,x_2)=\sum_{i=1}^\infty \sum_{j=1}^\infty P_F(\mathbf{Y}_i\in(x_1+i-1,x_1+i]\times(x_2+j-1,x_2+j])$. }
\end{figure}

\section{Estimation of the density function}\label{estimation}

 In the previous chapter we have derived inversion formulas that express the density $f$ in terms of the density $g$ of the observations. Now we can use an estimator  of $g$, for which we have observations, to estimate $f$. For an arbitrary density that is not of the form (\ref{convunbi}), the inversions will in general not yield distribution functions or densities, nor will they coincide. This typically happens if we estimate $g$.

We use kernel smoothing  but of course other estimators can be used as well. Let us introduce a bivariate  kernel density estimator with bivariate kernel function  $\mathbf{w}$ and bandwidth  $h>0$. The estimator $g_{nh}$ of $g$ is given by
\begin{equation}\label{kernel}
    g_{nh}(x_1,x_2)=\frac{1}{nh^2}\sum_{k=1}^n   \mathbf{w}\bigg(\frac{x_1-X_{k1}}{h},\frac{x_2-X_{k2}}{h}\bigg).
\end{equation}
Usually, $\mathbf{w}$ is chosen to be a bivariate probability density function. This way  it is ensured that $g_{nh}$ is also a density. See for instance Silverman (1986) and Wand and Jones (1995).

\bigskip

We impose the following condition on the kernel function.

\medskip

\noindent{\bf Condition $W$}

{\em The function $\mathbf{w}$ is a probability density function on
$\R^2$ with support $[-1,1]\times[-1,1]$. Furthermore, we will use a
product kernel $\mathbf{w}(u_1,u_2)=w_1(u_1)w_2(u_2)$, where
$w_i(u_i)$, with $i\in\{1,2\}$, denotes a continuously
differentiable univariate symmetric probability density function.}

\bigskip

We now substitute the kernel estimator in the four inversion
formulas of Lemma \ref{Inversion}. We  derive the estimator
$f^{++}_{nh}(x_1,x_2)$ as follows. The other three estimators follow
similarly. Define $w'_i(u):=\frac{d}{du}w_i(u), i=1,2$. Lemma
\ref{Inversion} in combination with $\frac{\partial^2}{\partial
x_1\partial x_2}F(x_1,x_2)=f(x_1,x_2)$ gives
\begin{align*}
    f_{nh}^{++}(x_1,x_2)&= \sum_{i=1}^\infty \sum_{j=1}^\infty \frac{\partial^2}{\partial x_1\partial x_2} g_{nh}(x_1+i,x_2+j)\\
    &= \sum_{i=1}^\infty \sum_{j=1}^\infty \bigg( \frac{\partial^2}{\partial x_1\partial x_2}\frac{1}{n}\sum_{k=1}^n \frac{1}{h^2} \mathbf{w}\bigg(\frac{x_1+i-X_{k1}}{h},\frac{x_2+j-X_{k2}}{h}\bigg)\bigg)\\
    &= \frac{1}{nh^4}\sum_{k=1}^n \sum_{i=1}^\infty \sum_{j=1}^\infty w'_1\bigg(\frac{x_1+i-X_{k1}}{h}\bigg) w'_2\bigg(\frac{x_2+j-X_{k2}}{h}\bigg).
\end{align*}
Note that, because of the bounded support of $\mathbf{w}$, the sum
is in fact a finite sum. In the last step we used the fact that
$\mathbf{w}$ is a product kernel, and thus
$\frac{\partial^2}{\partial u_1\partial
u_2}\mathbf{w}(u_1,u_2)=w'_1(u_1)w'_2(u_2)$.

The four kernel estimators of the density are given by
\begin{align*}
&f_{nh}^{--}(x_1,x_2) = \frac{1}{nh^4} \sum_{k=1}^n \sum_{i=0}^\infty \sum_{j=0}^\infty w'_1\Big(\frac{x_1-i-X_{k1}}{h}\Big) w'_2\Big(\frac{x_2-j-X_{k2}}{h}\Big),\\
&f_{nh}^{-+}(x_1,x_2) = -\frac{1}{nh^4} \sum_{k=1}^n \sum_{i=0}^\infty \sum_{j=1}^\infty w'_1\Big(\frac{x_1-i-X_{k1}}{h}\Big) w'_2\Big(\frac{x_2+j-X_{k2}}{h}\Big),\\
&f_{nh}^{+-}(x_1,x_2) = -\frac{1}{nh^4} \sum_{k=1}^n \sum_{i=1}^\infty \sum_{j=0}^\infty w'_1\Big(\frac{x_1+i-X_{k1}}{h}\Big) w'_2\Big(\frac{x_2-j-X_{k2}}{h}\Big),\\
&f_{nh}^{++}(x_1,x_2) = \frac{1}{nh^4} \sum_{k=1}^n
\sum_{i=1}^\infty \sum_{j=1}^\infty
w'_1\Big(\frac{x_1+i-X_{k1}}{h}\Big)
w'_2\Big(\frac{x_2+j-X_{k2}}{h}\Big).
\end{align*}

Next we introduce   a convex combination of the four previous estimators. Write
\begin{equation}
    f_{nh}^{(t)}(x_1,x_2)=t_1 f^{--}_{nh}(x_1,x_2)+t_2 f^{-+}_{nh}(x_1,x_2)+t_3 f^{+-}_{nh}(x_1,x_2)+t_4 f^{++}_{nh}(x_1,x_2),
\end{equation}
where $t=(t_1,t_2,t_3,t_4)$ and $t_1+t_2+t_3+t_4=1$. For suitable choices of $t_1,t_2,t_3,t_4$ this combination will  turn  out to have better properties than any of the estimators separately. Notice that when we set $t_1$, $t_2$, $t_3$, or $t_4$ equal to one and the others equal to zero, we   get   results for $f^{--}_{nh}$,$f^{-+}_{nh}$, $f^{+-}_{nh}$, or $f^{++}_{nh}$ individually.

\begin{thm}\label{mainthm1}
Assume that Condition $W$ is satisfied, that $f$ is bounded, and that\\ $lim_{x_1\to\pm\infty}f(x_1,x_2)=lim_{x_1\to\pm\infty}f(x_1,x_2)=0$. If $f$ is twice continuously differentiable on a neighborhood of $x=(x_1,x_2)$ then, as $n\to\infty, h\to 0, nh \to\infty$, we have
\begin{equation}\label{expectation}
    \ex f_{nh}^{(t)}(x_1,x_2)=f(x_1,x_2)+\frac{1}{2}h^2\Big(\int_{-\infty}^\infty z^2w_1(z)dz f_{11}(x_1,x_2)+\int_{-\infty}^\infty z^2w_2(z)dzf_{22}(x_1,x_2)\Big)+o(h^2).
\end{equation}

\noindent Furthermore, as $n\to\infty, h\to 0, nh \to\infty$, we have
\begin{equation}\label{variance1}
    \Var(f_{nh}^{(t)}(x_1,x_2))=\frac{1}{nh^6}B(x_1,x_2,t_1,t_2,t_3,t_4)\int_{-1}^1
    w'_1(z)^2dz
\int_{-1}^1w'_2(z)^2 dz+o(n^{-1}h^{-6})
\end{equation}
where
\begin{equation}\label{weights}
    B(x_1,x_2,t_1,t_2,t_3,t_4)=(t_1^2F^{--}+t_2^2F^{-+}+t_3^2F^{+-}+t_4^2F^{++})(x_1,x_2).
\end{equation}
\end{thm}
In the proof of the theorem we will see that the expectation of
$f_{nh}^{(t)}(x_1,x_2)$ is the same whatever convex combination we
choose for. Lemma \ref{optweights} gives the weights that minimize
the leading term in the variance (\ref{variance1}).

\begin{lem}\label{optweights}
Assume that $(x_1,x_2)$ is an interior point of the support of $f$. The weights $t_1$, $t_2$, $t_3$ and $t_4$, with $t_1+t_2+t_3+t_4=1$, that minimize the leading term in the variance (\ref{variance1}), are denoted by $\bar t_1(x_1,x_2)$, $\bar t_2(x_1,x_2)$, $\bar t_3(x_1,x_2)$ and $\bar t_4(x_1,x_2)$ and they are equal to
\begin{align*}
&\bar t_1(x_1,x_2)=F^{-+,+-,++}(x_1,x_2)A(x_1,x_2),\\
&\bar t_2(x_1,x_2)=F^{--,+-,++}(x_1,x_2)A(x_1,x_2),\\
&\bar t_3(x_1,x_2)=F^{--,-+,++}(x_1,x_2)A(x_1,x_2),\\
&\bar t_4(x_1,x_2)=F^{--,-+,+-}(x_1,x_2)A(x_1,x_2).
\end{align*}
The resulting variance of this optimal convex combination is then equal to
\begin{equation}
\Var(f_{nh}(x_1,x_2)) =A(x_1,x_2)C(x_1,x_2)\frac{1}{nh^6}\int_{-1}^1
    w'_1(z)^2dz
\int_{-1}^1w'_2(z)^2 dz+o(n^{-1}h^{-6}),
\end{equation}
Here
\begin{equation}
A(x_1,x_2):=(F^{-+,+-,++}+F^{--,+-,++}+F^{--,-+,++}+F^{--,-+,+-})^{-1}(x_1,x_2).
\end{equation}
where, for $a_1,a_2,b_1,b_2,c_1,c_2\in\{-,+\}$,
\begin{equation}
F^{a_1a_2,b_1b_2,c_1c_2}(x_1,x_2):=F^{a_1a_2}(x_1,x_2)F^{b_1b_2}(x_1,x_2)F^{c_1c_2}(x_1,x_2),
\end{equation}
and
\begin{equation}
C(x_1,x_2):=F^{--}(x_1,x_2)F^{-+}(x_1,x_2)F^{+-}(x_1,x_2)F^{++}(x_1,x_2).
\end{equation}
\end{lem}

\noindent{\bf Proof}

First note that the weights are well defined since the fact that $(x_1,x_2)$ is an interior point of the support of $f$ implies that $F^{--}(x_1,x_2), F^{-+}(x_1,x_2), F^{+-}(x_1,x_2)$ and $F^{++}(x_1,x_2)$ are strictly positive. The lower bound now follows from  Lemma \ref{inequality} in Chapter \ref{proofs}. \hfill$\Box$

\medskip

Note that in general, of course, we do not know $F$. However, in Section \ref{estweights} we show that we can estimate $F^{--}(x_1,x_2)$, $F^{-+}(x_1,x_2)$, $F^{+-}(x_1,x_2)$, and $F^{++}(x_1,x_2)$, again using the inversion formulas of Theorem \ref{Inversion}. This will lead to estimates of the optimal weights.
We then   prove that the estimator with estimated weights shares the properties of Theorem
\ref{mainthm1} with the optimal weights.

\section{The final estimator with estimated optimal weights}\label{estweights}

Let us write $\hat t_n(x_1,x_2)=(\hat t_{n1}(x_1,x_2), \ldots, \hat t_{n4}(x_1,x_2))$ for a vector
of estimated weights. The next theorem shows that under some conditions on these estimators the   limit behaviour
of $ f_{nh}^{(\hat t_n)}(x_1,x_2)$  resembles the optimal limit behaviour of the estimator $ f_{nh}^{(\bar t)}(x_1,x_2)$.
\begin{thm}\label{mainthmest}
Assume that Condition $W$ is satisfied, that $f$ is bounded, and that\\ $lim_{x_1\to\pm\infty}f(x_1,x_2)=lim_{x_1\to\pm\infty}f(x_1,x_2)=0$.

Assume for $i=1,\ldots ,4$,
\begin{equation}\label{c1}
\ex (\hat t_{ni}(x_1,x_2) - \bar t_i(x_1,x_2))^2=o(nh^{10}).
\end{equation}
If $f$ is twice continuously differentiable on a neighborhood of $x=(x_1,x_2)$ then, as $n\to\infty, h\to 0, nh \to\infty$,
we have
\begin{equation}\label{expectation2}
    \ex f_{nh}^{(\hat t_n)}(x_1,x_2)=f(x_1,x_2)+\frac{1}{2}h^2\Big(\int_{-\infty}^\infty z^2w_1(z)dz f_{11}(x_1,x_2)+\int_{-\infty}^\infty z^2w_2(z)dzf_{22}(x_1,x_2)\Big)+o(h^2).
\end{equation}

Assume for $i=1,\ldots ,4$,
\begin{equation}\label{c2}
\ex (\hat t_{ni}(x_1,x_2) - \bar t_i(x_1,x_2))^4=o(1).
\end{equation}
\noindent Then, as $n\to\infty, h\to 0, nh \to\infty$, we have
\begin{equation}
\Var(f_{nh}^{(\hat t_n)}(x_1,x_2))
= \frac{1}{nh^6}\,\sigma(x_1,x_2)^2+o(n^{-1}h^{-6}),
\end{equation}
where, with the notation of Lemma \ref{optweights}, $\sigma(x_1,x_2)^2$ is defined by
\begin{equation}
\sigma(x_1,x_2)^2 =A(x_1,x_2)C(x_1,x_2) \int_{-1}^1
    w'_1(z_1)^2dz_1
\int_{-1}^1w'_2(z_2)^2 dz_2.
\end{equation}
Assume for $i=1,\ldots ,4$,
\begin{equation}\label{c3}
\ex (\hat t_{ni}(x_1,x_2) - \bar t_i(x_1,x_2))^2=o(1).
\end{equation} Then the estimator is  asymptotically normally distributed. We have,
,as $n\to\infty, h\to 0, nh \to\infty$,
\begin{equation}
\sqrt{n}h^3 \Big(f_{nh}^{(\hat t_n)}(x_1,x_2) - \ex f_{nh}^{(\hat t_n)}(x_1,x_2)\Big)
\convd N(0,\sigma(x_1,x_2)^2).
\end{equation}
\end{thm}

\bigskip

Let us next construct suitable estimators of the weights based on
the estimators of $F^{--} ,F^{-+},$ $F^{+-}$ and $F^{++}$. As in
estimation of the density we can plug in (\ref{kernel}) into the
inversion  formulas for $F$ in Lemma \ref{Inversion} and get kernel
estimators of $F^{--}(x_1,x_2),F^{-+}(x_1,x_2),F^{+-}(x_1,x_2)$ and
$F^{++}(x_1,x_2)$. We get four estimators, given by
\begin{align}\label{Fnhestimators}
                F_{nh}^{--}(x_1,x_2) = \frac{1}{nh^2} \sum_{k=1}^n \sum_{i=0}^\infty \sum_{j=0}^\infty w_1\left(\frac{x_1-i-X_{k1}}{h}\right) w_2\left(\frac{x_2-j-X_{k2}}{h}\right),\nonumber\\
        F_{nh}^{-+}(x_1,x_2) = \frac{1}{nh^2} \sum_{k=1}^n \sum_{i=0}^\infty \sum_{j=1}^\infty w_1\left(\frac{x_1-i-X_{k1}}{h}\right) w_2\left(\frac{x_2+j-X_{k2}}{h}\right),\nonumber\\
        F_{nh}^{+-}(x_1,x_2) = \frac{1}{nh^2} \sum_{k=1}^n \sum_{i=1}^\infty \sum_{j=0}^\infty w_1\left(\frac{x_1+i-X_{k1}}{h}\right) w_2\left(\frac{x_2-j-X_{k2}}{h}\right),\nonumber\\
        F_{nh}^{++}(x_1,x_2) = \frac{1}{nh^2} \sum_{k=1}^n \sum_{i=1}^\infty \sum_{j=1}^\infty w_1\left(\frac{x_1+i-X_{k1}}{h}\right) w_2\left(\frac{x_2+j-X_{k2}}{h}\right).
    \end{align}

\bigskip

The following theorem  establishes the asymptotic bias and variance
of these four estimators. In the sequel we adopt the notation
 $F^{--}_{11}=\frac{\partial^2
F^{--}(x_1,x_2)}{\partial x_1^2}$ and $F^{--}_{12}=\frac{\partial^2
F(x_1,x_2)}{\partial x_1\partial x_2}$, etc., also for the density
$f$. The proof is very similar to the proof of Theorem \ref{mainthm1} and is therefore omitted. See
Bene\v{s}ov\'{a} et al. (2011) for a complete proof.

\begin{thm}\label{mainthm2}
Assume that Condition $W$ is satisfied. Then, as $n\to\infty, h\to
0, nh \to\infty$ we have
\begin{eqnarray*}\label{allexp2dim}
    \ex F_{nh}^{--}(x_1,x_2)=F^{--}(x_1,x_2)+ \frac{1}{2}h^2\Big(\int_{-\infty}^\infty z^2w_1(z)dz F^{--}_{11}(x_1,x_2)+\int_{-\infty}^\infty z^2w_2(z)dzF^{--}_{22}(x_1,x_2)\Big)+o(h^2), \\
    \ex F_{nh}^{-+}(x_1,x_2)=F^{-+}(x_1,x_2)+ \frac{1}{2}h^2\Big(\int_{-\infty}^\infty z^2w_1(z)dz F^{-+}_{11}(x_1,x_2)+\int_{-\infty}^\infty z^2w_2(z)dzF^{-+}_{22}(x_1,x_2)\Big)+o(h^2),  \\
    \ex F_{nh}^{+-}(x_1,x_2)=F^{+-}(x_1,x_2) +  \frac{1}{2}h^2\Big(\int_{-\infty}^\infty z^2w_1(z)dz F^{+-}_{11}(x_1,x_2)+\int_{-\infty}^\infty z^2w_2(z)dzF^{+-}_{22}(x_1,x_2)\Big)+o(h^2), \\
    \ex F_{nh}^{++}(x_1,x_2)=F^{++}(x_1,x_2) +  \frac{1}{2}h^2\Big(\int_{-\infty}^\infty z^2w_1(z)dz F^{++}_{11}(x_1,x_2)+\int_{-\infty}^\infty z^2w_2(z)dzF^{++}_{22}(x_1,x_2)\Big)+o(h^2).
\end{eqnarray*}
 For the variances we have
\begin{eqnarray*}\label{allvar2dim}
\Var(F^{--}_{nh}(x_1,x_2))=F^{--}(x_1,x_2)\frac{1}{nh^2} \int_{-1}^{1} w_1^2(z ) dz  \int_{-1}^{1} w_2^2(z ) dz + o\left(\frac{1}{nh^{2}}\right),\\
\Var(F^{-+}_{nh}(x_1,x_2))=F^{-+}(x_1,x_2)\frac{1}{nh^2} \int_{-1}^{1} w_1^2(z ) dz  \int_{-1}^{1} w_2^2(z ) dz + o\left(\frac{1}{nh^{2}}\right),\\
\Var(F^{+-}_{nh}(x_1,x_2))=F^{+-}(x_1,x_2)\frac{1}{nh^2} \int_{-1}^{1} w_1^2(z ) dz  \int_{-1}^{1} w_2^2(z ) dz + o\left(\frac{1}{nh^{2}}\right),\\
\Var(F^{++}_{nh}(x_1,x_2))=F^{++}(x_1,x_2)\frac{1}{nh^2}
\int_{-1}^{1} w_1^2(z ) dz  \int_{-1}^{1} w_2^2(z ) dz +
o\left(\frac{1}{nh^{2}}\right).
\end{eqnarray*}
\end{thm}
For the proof of this theorem see Chapter \ref{proofs}.

\medskip

Next we write the optimal weights of Lemma \ref{optweights} in terms of functions
$\tilde t_i$ defined by
$$\bar t_i(x_1,x_2)=\tilde t_i( F^{--}(x_1,x_2),  F ^{-+}(x_1,x_2),  F ^{+-}(x_1,x_2),
 F ^{++}(x_1,x_2)), \quad i=1,\ldots,4.
$$
Let $(\epsilon_n)$ denote a sequence of numbers with
$0< \epsilon_n<1$ and $\epsilon_n\to 0$ as $n\to\infty$.
Then define truncated versions of the estimators $F_{nh}^{--}(x_1,x_2), F_{nh}^{-+}(x_1,x_2), F_{nh}^{+-}(x_1,x_2),F_{nh}^{+-}(x_1,x_2)$ and $F_{nh}^{++}(x_1,x_2)$ by
\begin{align*}
&{\tilde F}_{nh}^{--}(x_1,x_2) = \min(\max(F _{nh}^{--}(x_1,x_2),\epsilon_n),1),\\
&{\tilde F}_{nh}^{-+}(x_1,x_2) =  \min(\max(F _{nh}^{-+}(x_1,x_2),\epsilon_n),1) ,\\
&{\tilde F}_{nh}^{+-}(x_1,x_2) =  \min(\max(F _{nh}^{+-}(x_1,x_2),\epsilon_n),1) ,\\
&{\tilde F}_{nh}^{++}(x_1,x_2) =  \min(\max(F _{nh}^{++}(x_1,x_2),\epsilon_n),1) .
\end{align*}
Since the bandwidth used in the estimators  of the weights can in general be different to the bandwidth $h$ used in the  estimator of  $f$, we will denote this bandwidth by $\tilde h$. We now obtain estimators of the weights by plugging in these estimators. We get
$$
\hat t_{ni}(x_1,x_2)=\tilde t_i({\tilde F}_{n\tilde h}^{--}(x_1,x_2), {\tilde F}_{n\tilde h}^{-+}(x_1,x_2), {\tilde F}_{n\tilde h}^{+-}(x_1,x_2)),
{\tilde F}_{n\tilde h}^{++}(x_1,x_2)), \quad i=1,\ldots,4.
$$
The next lemma shows that these estimators, with a suitable bandwidth, can be used to estimate the optimal weights without disturbing the asymptotics of Theorem \ref{mainthm1}.
\begin{lem}\label{checkest}
If $h \gg n^{-1/6}$, $\epsilon_n=1/\log n$,  and if  we use a bandwidth $\tilde h$ of the form $\tilde h=cn^{-1/6}$, where $c$ is a constant,
then the estimators
$$
\hat t_{ni}(x_1,x_2)=\tilde t_i({\tilde F}_{n\tilde
h}^{--}(x_1,x_2), {\tilde F}_{n\tilde h}^{-+}(x_1,x_2), {\tilde
F}_{n\tilde h}^{+-}(x_1,x_2), {\tilde F}_{n\tilde h}^{++}(x_1,x_2))
$$
 satisfy (\ref{c1}), (\ref{c2}) and (\ref{c3}).
\end{lem}

\begin{rem}
If we compare the performance of our final estimator with estimated
optimal weights to the performance of the four individual estimators
then we see that the first order of the expectation is the same. The
variance of the combined estimator contains the term $C(x_1,x_2)$
which is equal to the product of $F^{--}(x_1,x_2), F^{-+}(x_1,x_2),
F^{+-}(x_1,x_2)$ and $F^{++}(x_1,x_2)$. This shows that the variance
is small along the edge of the support of $f$. By Theorem
\ref{mainthm1} the variance of, for instance, $f^{--}_{nh}(x_1,x_2)$
is proportional to $F^{--}(x_1,x_2)$. So this estimator will perform
better in the lower left of the support of $f$ than it will in the
other part. By using the estimated optimal convex combination the
worse behavior of the four individual estimators in certain areas is
reduced.
\end{rem}
\begin{rem}
Since in the theorems we use a bivariate kernel function $\mathbf{w}$ which is the product of two different univariate density functions $w_1$ and $w_2$, in fact we allow different bandwidths for the two coordinates, provided the bandwidths are of the same order. Writing $h_1= h,h_2=ch, w_1=w$ and $w_2=w(\cdot/c)/c$, for some $c>0$, and writing $f_{nh_1h_2}^{(t)}$ for the resulting estimator,
we get the following leading terms in the expansions of its bias and variance in Theorem \ref{mainthm1},
\begin{equation}\label{expectationdiff}
    \frac{1}{2}  \int_{-\infty}^\infty z^2w(z)dz \Big(h_1^2 f_{11}(x_1,x_2)+h_2^2
    f_{22}(x_1,x_2)\Big)
\end{equation}
and
\begin{equation}\label{variance1diff}
\frac{1}{nh_1^3h_2^3}B(x_1,x_2,t_1,t_2,t_3,t_4)\Big(\int_{-1}^1
    w'(z)^2dz\Big)^2.
\end{equation}
The subsequent theorems can be likewise adapted to different bandwidths.
\end{rem}
\begin{rem}
If we minimize the pointwise asymptotic mean squared error of
$f_{nh}^{(\hat t_n)}(x_1,x_2)$ and thus balance its asymptotic
squared bias and its asymptotic variance
 given by Theorem \ref{mainthmest} then we see that the optimal bandwidth is of order $n^{-1/10}$. The corresponding mean squared error is then equal to $n^{-2/5}$. This of course raises the problem of
bandwidth selection which, important though as it is for applications, we will not pursue here.
\end{rem}
\begin{rem}
In the proofs we see that the bias of our final estimator is asymptotically of the same form as the bias of a bivariate kernel density estimator based on direct observations. That means that, if the smoothness assumptions on the density $f$ are strengthened, bias reduction techniques, such as for instance higher order kernels or even super kernels, can be used to increase the rate of convergence.
\end{rem}
\begin{rem}
The construction as presented here for bivariate data can in principle also be done for arbitrary dimension $d$. For dimension one we have to combine two inversion formulas as shown in Van Es (2011). In the present paper, for dimension two, we combine four inversion formulas, and for arbitrary dimension $d$ combination of $2^d$ inversions has to be accomplished. Of course the complexity of the estimator will increase rapidly with growing dimension.
\end{rem}

\section{Simulated examples}\label{simulation}

To illustrate the estimator we have simulated two examples. In the first example the density $f$ is unimodal.
In the second example $f$ is a mixture of two unimodal bivariate densities, rendering it bimodal. In the first example $f$ is concentrated on the square $[0.25,1.75]\times [0.25,1.75]$. In the second example $f$ is concentrated on the square $[0.2,1.8]\times [0.2,1.8]$. This means that both deconvolution problems  are not at all trivial.

To speed up computations we have followed the bivariate binning technique as advised in Wand (1994). For the $x$ and $y$ coordinates we have chosen for a grid of 500 points between -1 and 4.
We have used a product kernel based on the so called biweight kernel given by
\begin{equation}
w_1(u)=w_2(u)=\frac{15}{16}\, (1-u^2)^2\I_{[-1,1]}(u).
\end{equation}

\begin{exam}\label{beta}{\rm
In our first example $f$ is the density of the random vector $(Y_1,Y_2)$, where $Y_1$ and $Y_2$ are two independent random variables that each have a certain shifted and rescaled beta distribution.
To be more specific $Y_i= 0.25 + 1.5 V_i, i=1,2$, where the $V_i$ are independent and both Beta(3,3) distributed.
We have simulated 1000 values so $n=1000$. The bandwidth $h$, chosen by hand, is equal to $0.5$.

The true density $f$ and its estimate are given in Figure  \ref{fig1} . The difference between the true density and the estimate is plotted in Figure \ref{fig2}. The right plot in Figure \ref{fig2} shows $f_{nh}^{+-}$. Clearly this estimate
is best in the $+-$ quadrant, as predicted by the theory.

\begin{figure}[h]
$$
\includegraphics[width=8cm]{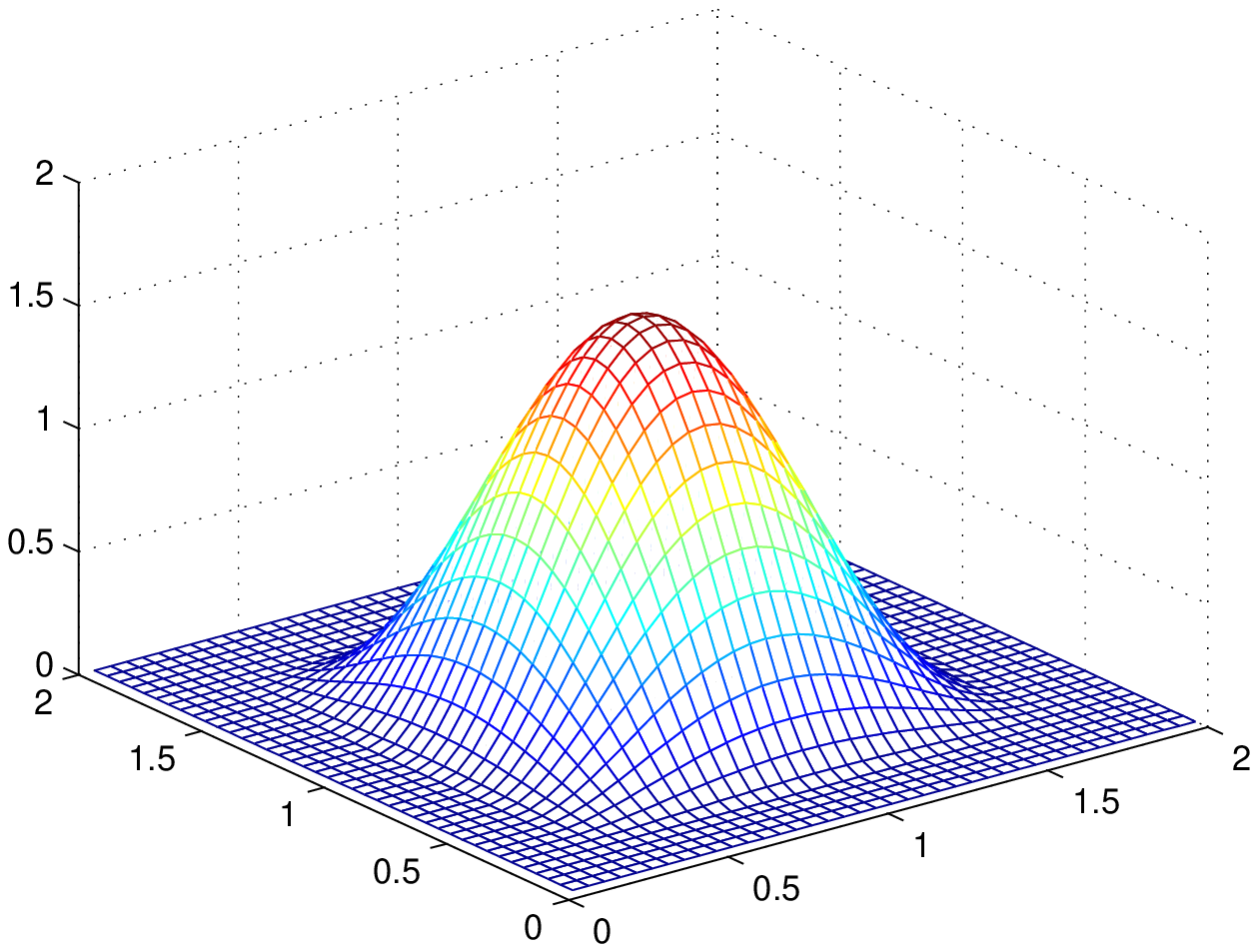}
\includegraphics[width=8cm]{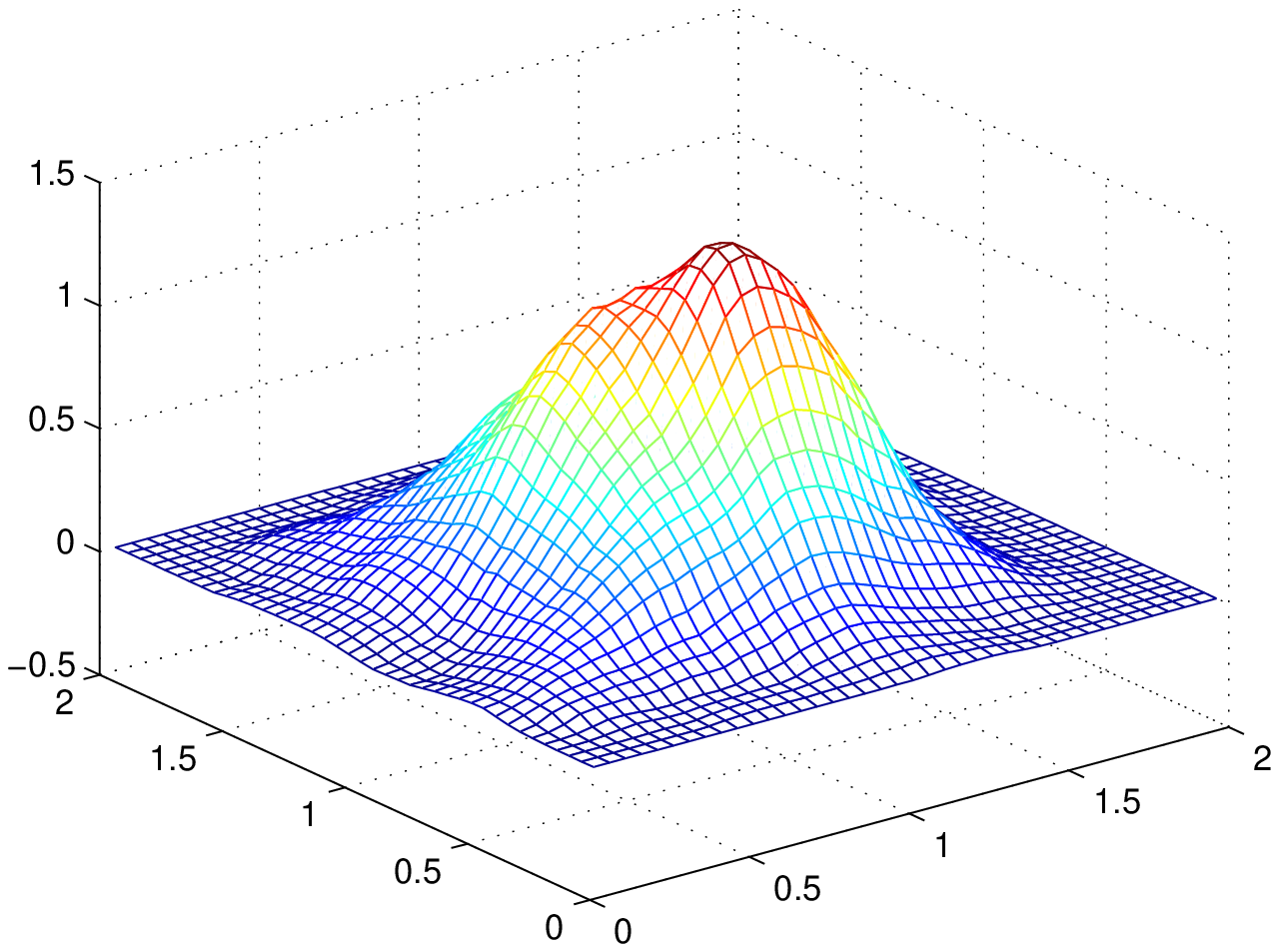}
$$
\caption[]{  Left: the true density. Right: the estimate.\label{fig1}}
\end{figure}
\begin{figure}[h]
$$
\includegraphics[width=8cm]{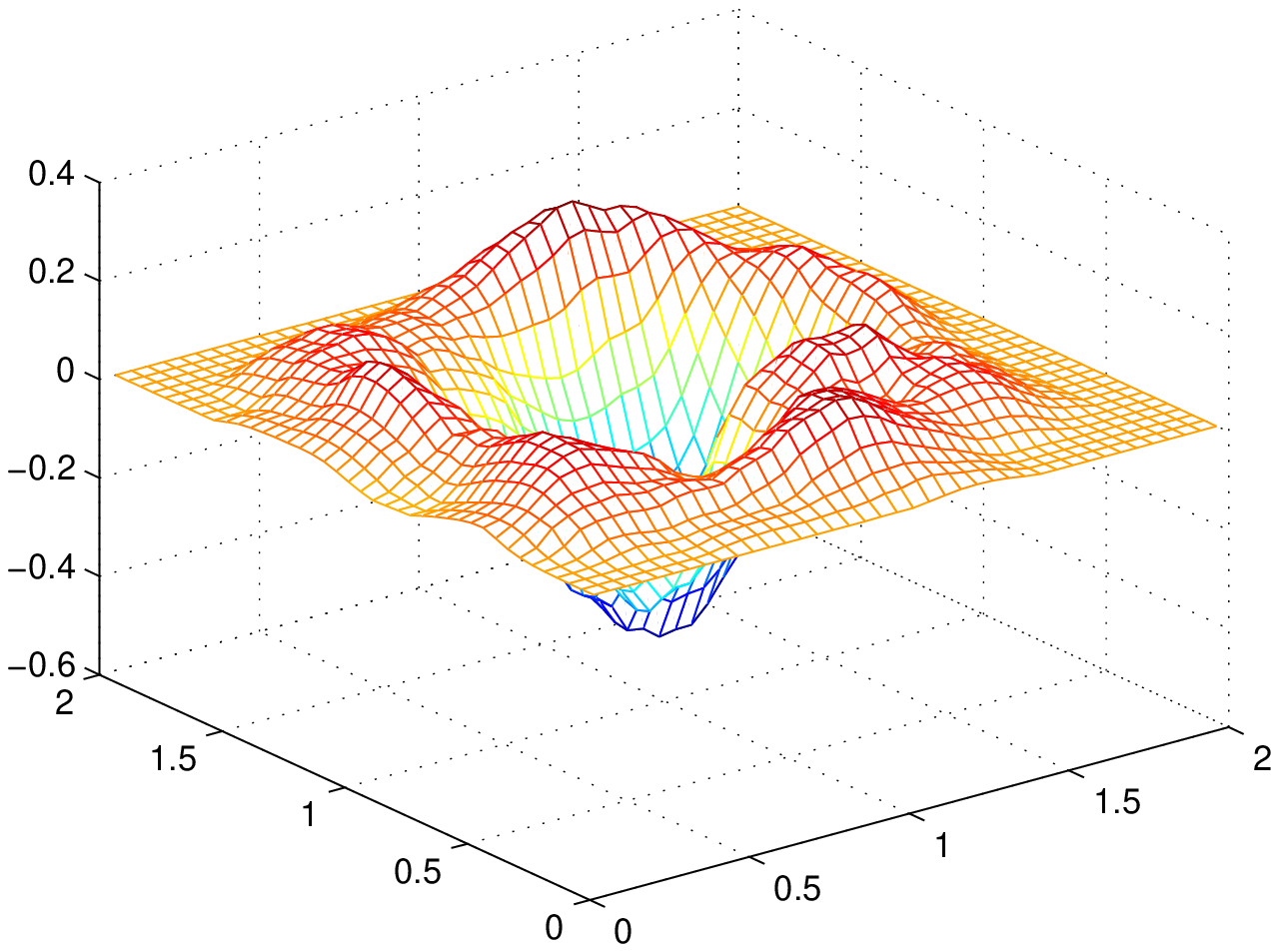}
\includegraphics[width=8cm]{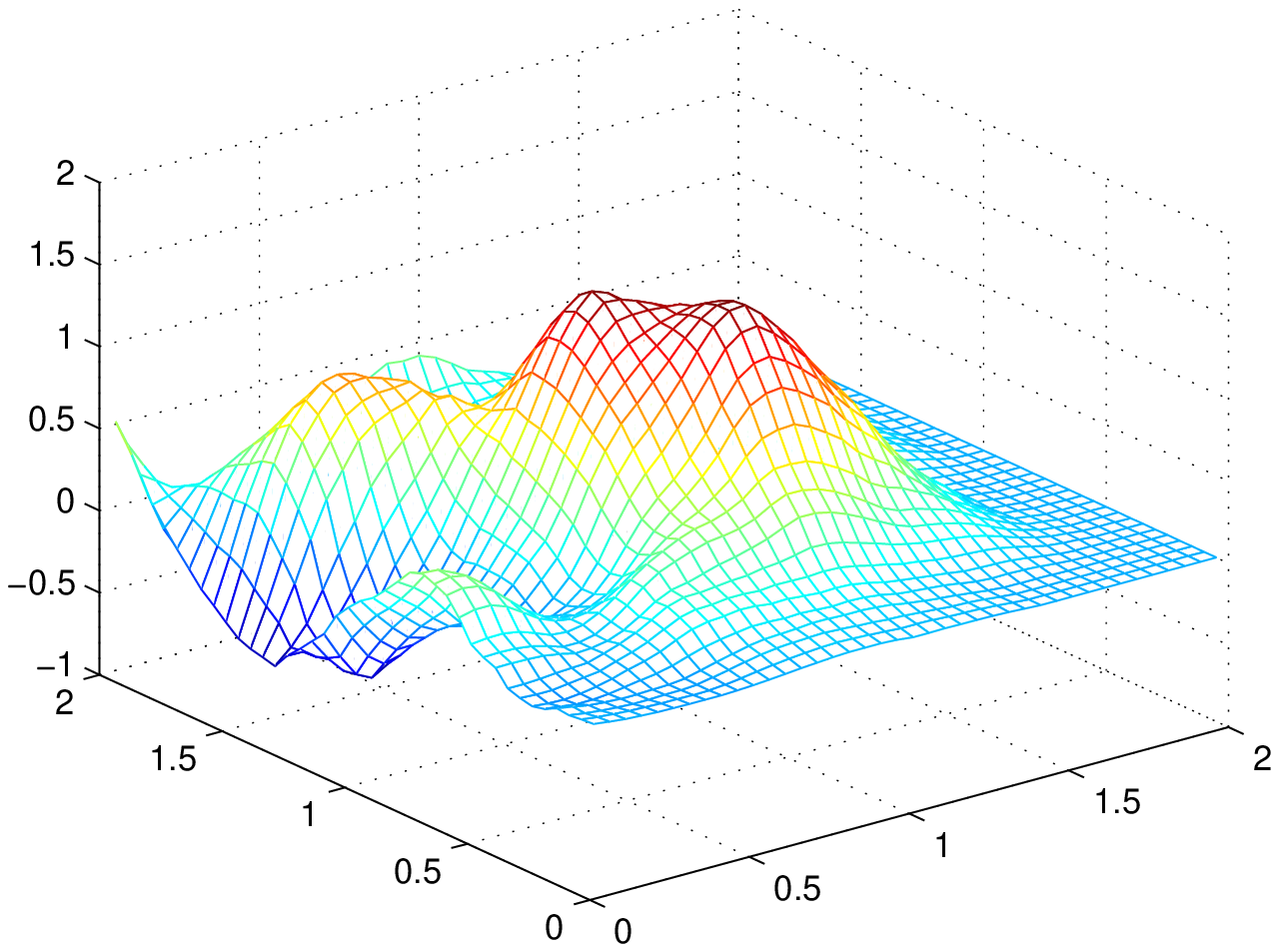}
$$
\caption[]{Left: the difference of the true density and the estimate. Right: $f_{nh}^{+-}$.\label{fig2} }
\end{figure}

}
\end{exam}

\begin{exam}{\rm
In our second example $f$ is the density of the random vector $(Y_1,Y_2)$, where $Y_1$ and $Y_2$ are dependent random variables with a bimodal distribution. The distribution of the vector is a mixture of two distributions like the one in Example \ref{beta}.
The values of the $Y$'s are generated as follows.  With $V_1$ and $V_2$ having the same distribution as in the previous example
the $Y$ values are given by
$$
\begin{pmatrix}
    Y_{1}\\
    Y_{2}
\end{pmatrix}
=
\left\{
\begin{array}{ll}
\begin{pmatrix}
     V_{1}+0.2\\
     V_{2}+0.8
\end{pmatrix}
&\mbox{, with probability 2/5,}
\\
&\\
\begin{pmatrix}
     V_{1}+0.8\\
    V_{2}+0.2
\end{pmatrix}
&\mbox{, with probability 3/5.}
\end{array}
\right.
$$
We have simulated 5000 values so $n=5000$. The bandwidth $h$, chosen by hand, is equal to $0.35$.

 The true density $f$ and its estimate are given in Figure  \ref{fig3} . The difference between the true density and the estimate is plotted in Figure \ref{fig4}. The right plot in Figure \ref{fig4} shows $f_{nh}^{-+}$. Clearly this estimate
is best in the $-+$ quadrant, as predicted by the theory.

\begin{figure}[h]
$$
\includegraphics[width=8cm]{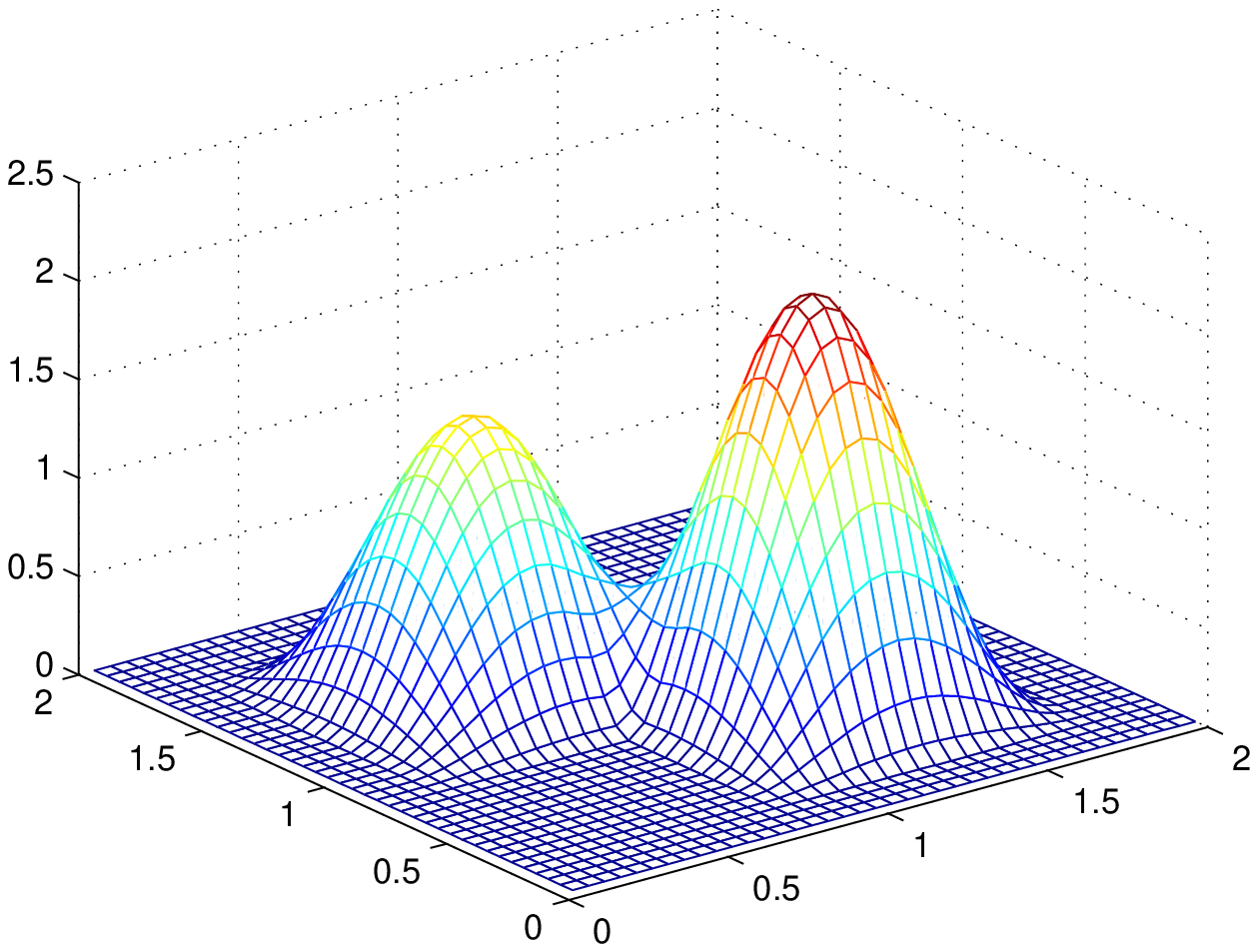}
\includegraphics[width=8cm]{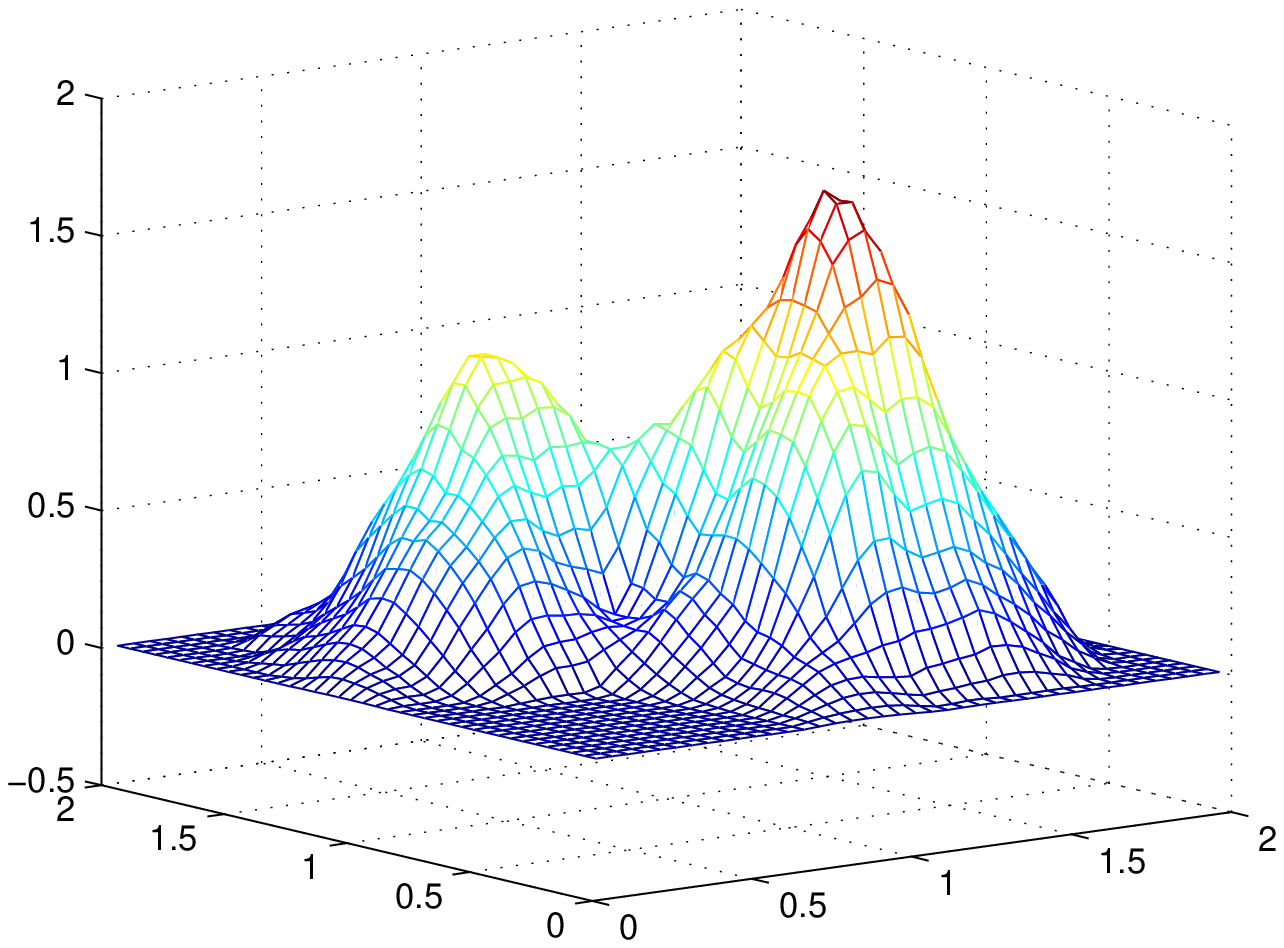}
$$
\caption[]{$n=5000, h=0.35$. Left: the true density. Right: the estimate.\label{fig3} }
\end{figure}
\begin{figure}[h]
$$
\includegraphics[width=8cm]{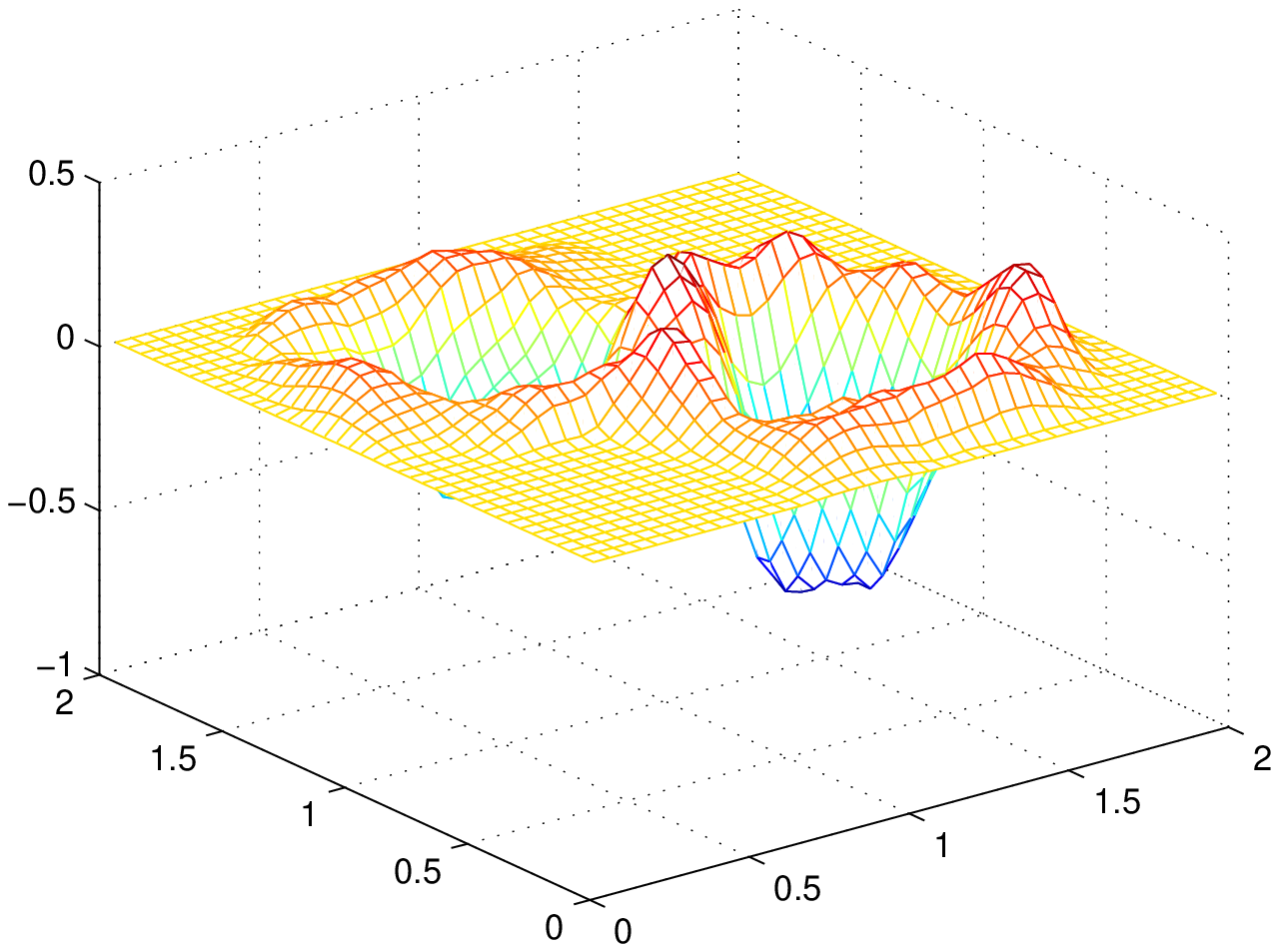}
\includegraphics[width=8cm]{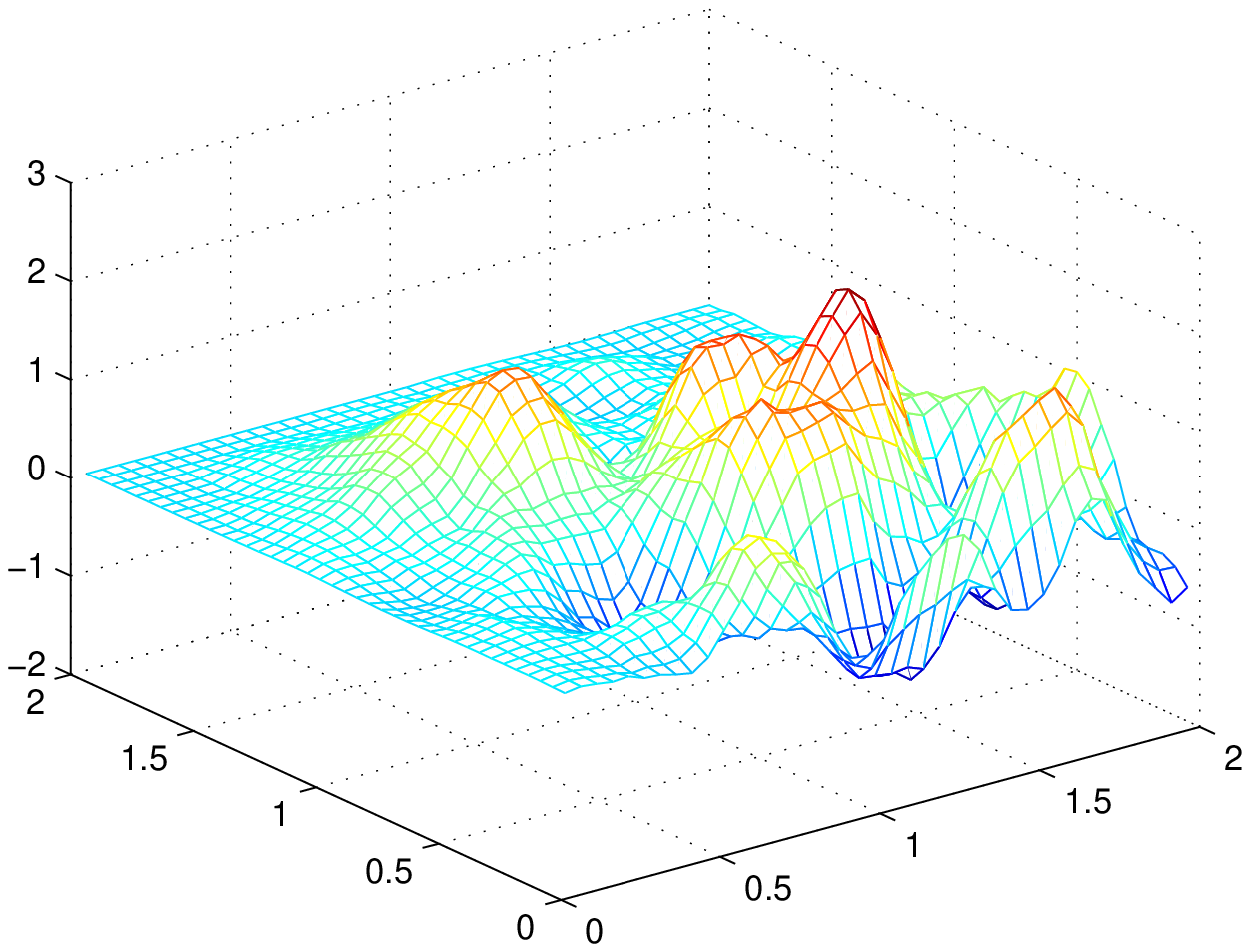}
$$
\caption[]{Left: the difference of the true density and the estimate. Right: $f_{nh}^{-+}$.\label{fig4}}
\end{figure}

}
\end{exam}

\section{Proofs}\label{proofs}

\subsection{Proof of Lemma \ref{Inversion}}

Let us first derive the inversion formulas for $F(x_1,x_2)$. We sum
$g(x_1-i,x_2) = F(x_1-i,x_2) - F(x_1-i,x_2-1) - F(x_1-i-1,x_2) +
F(x_1-i-1,x_2-1)$ over the first coordinate to obtain two telescopic
sums. Thus we get
\begin{align}\label{telescope1}
    &\sum_{i=0}^\infty g(x_1-i,x_2)\nonumber\\
    &= \sum_{i=0}^\infty\{F(x_1-i,x_2) - F(x_1-i,x_2-1) - F(x_1-i-1,x_2) + F(x_1-i-1,x_2-1)\}\nonumber\\
    &= \sum_{i=0}^\infty\{F(x_1-i,x_2) - F(x_1-i-1,x_2)\} - \sum_{i=0}^\infty\{F(x_1-i,x_2-1) - F(x_1-i-1,x_2-1)\}\nonumber\\
    &= F(x_1,x_2) - F(x_1,x_2-1).
\end{align}
Here we used that $\lim_{i \to \infty}F(x_1-i,x_2)=\lim_{i \to \infty}F(x_1-i,x_2-1)=0$, for $F$ is a bivariate distribution function. Next, we sum over the second coordinate. Because we also have $\lim_{j \to \infty}F(x_1,x_2-j)=0$, we get
\begin{equation}\label{telescope2}
     \sum_{j=0}^\infty \sum_{i=0}^\infty g(x_1-i,x_2-j) = \sum_{j=0}^\infty \{F(x_1,x_2-j) - F(x_1,x_2-j-1)\} = F(x_1,x_2).
\end{equation}

\noindent Because the terms are nonnegative, the order of summation can be interchanged and we have shown (\ref {--}). Thus we have found an expression for the unobservable probability distribution function $F$ in terms of the observable density function $g$.

Above, we iterated over $-i$, so now let us determine what happens when we iterate over $+i$. First, we write $g(x_1+i,x_2)$ as
\begin{equation}\label{invplus}
    g(x_1+i,x_2) = F(x_1+i,x_2) - F(x_1+i,x_2-1) - F(x_1+i-1,x_2) + F(x_1+i-1,x_2-1).
\end{equation}
Secondly, we take the sum over the first coordinate. Again we get two telescopic sums. Note that $\lim_{i\to\infty}F(x_1+i,x_2)=F_{Y_2}(x_2)$ and $\lim_{i\to\infty}F(x_1+i,x_2-1)=F_{Y_2}(x_2-1)$, so we get
\begin{align}\label{invplus2}
    &\sum_{i=1}^\infty g(x_1+i,x_2)\nonumber\\
    &=\sum_{i=1}^\infty\{F(x_1+i,x_2) - F(x_1+i,x_2-1) - F(x_1+i-1,x_2) + F(x_1+i-1,x_2-1)\}\nonumber\\
    &=\sum_{i=1}^\infty\{F(x_1+i,x_2) - F(x_1+i-1,x_2)\} + \sum_{i=1}^\infty\{F(x_1+i-1,x_2-1) - F(x_1+i,x_2-1)\}\nonumber\\
    &=F_{Y_2}(x_2) - F(x_1,x_2) + F(x_1,x_2-1) - F_{Y_2}(x_2-1).
\end{align}
Thirdly, we sum over the second coordinate. Because $\lim_{j \to \infty}F_{Y_2}(x_2-j)=0$, this results in
\begin{align}
    &\sum_{j=0}^\infty\sum_{i=1}^\infty g(x_1+i,x_2-j)\nonumber\\
    &=\sum_{j=0}^\infty\{F_{Y_2}(x_2-j) - F(x_1,x_2-j) + F(x_1,x_2-j-1) - F_{Y_2}(x_2-j-1)\}\nonumber\\
    &=\sum_{j=0}^\infty\{F_{Y_2}(x_2-j) - F_{Y_2}(x_2-j-1)\} - \sum_{j=0}^\infty\{F(x_1,x_2-j) - F(x_1,x_2-j-1)\}\nonumber\\
    &=F_{Y_2}(x_2) - F(x_1,x_2)=F^{+-}(x_1,x_2).
\end{align}
Again, we can interchange the sums and we have shown (\ref{+-}). In similar fashion we can derive (\ref{-+}).

The last formula to recover $F(x_1,x_2)$ can be derived as follows. We begin with
\begin{equation}
g(x_1+1,x_2+1)=F(x_1,x_2)-F(x_1,x_2+1)-F(x_1+1,x_2)+F(x_1+1,x_2+1).
\end{equation}
Now sum over the first coordinate to obtain
\begin{equation}
    \sum_{i=1}^\infty g(x_1+i,x_2+1) = F(x_1,x_2) - F(x_1,x_2+1) - F_{Y_2}(x_2) + F_{Y_2}(x_2+1).
\end{equation}
Summing over the second coordinate we get
\begin{equation}
    \sum_{j=1}^\infty \sum_{i=1}^\infty g(x_1+i,x_2+j) = F(x_1,x_2) - F_{Y_1}(x_1) - F_{Y_2}(x_2) + 1=F^{++}(x_1,x_2).
\end{equation}
Changing the order of summation again, we obtain $(\ref{++})$.

The four inversion formulas for $f$ are derived in a similar fashion.  From (\ref{convunbi}) we have
$$\frac{\partial^2}{\partial x_1\partial x_2}g(x_1,x_2)=f(x_1,x_2) - f(x_1,x_2-1) - f(x_1-1,x_2) + f(x_1-1,x_2-1).
$$
Now, following equations (\ref{telescope1}) and (\ref{telescope2}), we obtain
\begin{equation}
    \sum_{i=0}^\infty \sum_{j=0}^\infty\frac{\partial^2}{\partial x_1\partial x_2}g(x_1-i,x_2-j)=f(x_1,x_2).
\end{equation}
Here we have used  $\lim_{x_1 \to -\infty} f(x_1,x_2)=0$ and $\lim_{x_2 \to -\infty} f(x_1,x_2)=0$.

The other three inversion formulas follow similarly.\hfill$\Box$

\subsection{Proof of Theorem \ref{mainthm1}}

First we consider the estimator $f_{nh}^{++}$. We have
\begin{align}\label{exp}
    \ex f_{nh}^{++}&(x_1,x_2)= \ex \bigg(\frac{1}{nh^4} \sum_{k=1}^n\sum_{i=1}^\infty \sum_{j=1}^\infty w'_1\bigg(\frac{x_1+i-X_{k1}}{h}\bigg) w'_2\bigg(\frac{x_2+j-X_{k2}}{h}\bigg)\bigg)\nonumber\\
    &= \frac{1}{h^4} \sum_{i=1}^\infty \sum_{j=1}^\infty \ex w'_1\bigg(\frac{x_1+i-X_{11}}{h}\bigg) w'_2\bigg(\frac{x_2+j-X_{12}}{h}\bigg)\nonumber\\
    &= \frac{1}{h^4} \sum_{i=1}^\infty \sum_{j=1}^\infty \int_{-\infty}^\infty \int_{-\infty}^\infty w'_1\Big(\frac{x_1+i-u_1}{h}\Big) w'_2\Big(\frac{x_2+j-u_2}{h}\Big)g(u_1,u_2)du_1du_2.
\end{align}
\medskip
Note that interchanging integrals and sums is allowed because
\begin{equation}\label{exp1}
    \frac{1}{h^4} \sum_{i=1}^\infty \sum_{j=1}^\infty \int_{-\infty}^\infty \int_{-\infty}^\infty \Big|w'_1\Big(\frac{x_1+i-u_1}{h}\Big)\Big| \Big|w'_2\Big(\frac{x_2+j-u_2}{h}\Big)\Big|g(u_1,u_2)du_1du_2<\infty.
\end{equation}
To check this, we first make the substitutions $v_1:=u_1-i$ and $v_2:=u_2-j$. Secondly, we interchange the sums and integrals again, which is allowed because the integrand is nonnegative (Fubini). We get
\begin{align}\label{exp2}
    &\frac{1}{h^4} \sum_{i=1}^\infty \sum_{j=1}^\infty \int_{-\infty}^\infty \int_{-\infty}^\infty \Big|w'_1\Big(\frac{x_1-v_1}{h}\Big)\Big| \Big|w'_2\Big(\frac{x_2-v_2}{h}\Big)\Big|g(v_1+i,v_2+j)dv_1dv_2\nonumber\\
    &=\frac{1}{h^4} \int_{-\infty}^\infty \int_{-\infty}^\infty\Big|w'_1\Big(\frac{x_1-v_1}{h}\Big)\Big|\Big|w'_2\Big(\frac{x_2-v_2}{h}\Big)\Big|\sum_{i=1}^\infty \sum_{j=1}^\infty g(v_1+i,v_2+j)dv_1dv_2.
\end{align}
Thirdly, noting that $F^{++}(v_1,v_2)=\sum_{i=1}^\infty \sum_{j=1}^\infty g(v_1+i,v_2+j)dv_1dv_2$ and that $F^{++}(v_1,v_2)\leq1$, we obtain
\begin{align}\label{exp25}
    &\frac{1}{h^4} \int_{-\infty}^\infty \int_{-\infty}^\infty\Big|w'_1\Big(\frac{x_1-v_1}{h}\Big)\Big|\Big|w'_2\Big(\frac{x_2-v_2}{h}\Big)\Big| F^{++}(v_1,v_2)dv_1dv_2\nonumber\\
    &\leq\frac{1}{h^4} \int_{-\infty}^\infty \int_{-\infty}^\infty\Big|w'_1\Big(\frac{x_1-v_1}{h}\Big)\Big|\Big|w'_2\Big(\frac{x_2-v_2}{h}\Big)\Big|dv_1dv_2<\infty.
\end{align}
Because $w_1'$ and  $w_2'$ are bounded functions, and have bounded
support, this integral is finite. Thus our use of Fubini's Theorem
is justified. Next we apply partial integration twice, yielding
\medskip
\begin{align*}
    \ex f_{nh}^{++}(&x_1,x_2)=\frac{1}{h^4} \sum_{i=1}^\infty \sum_{j=1}^\infty \int_{-\infty}^\infty w'_2\Big(\frac{x_2+j-u_2}{h}\Big) \Bigg(\int_{-\infty}^\infty w'_1\Big(\frac{x_1+i-u_1}{h}\Big) g(u_1,u_2)du_1\Bigg)du_2\\
    &= -\frac{1}{h^3} \sum_{i=1}^\infty \sum_{j=1}^\infty \int_{-\infty}^\infty w'_2\Big(\frac{x_2+j-u_2}{h}\Big)\Bigg(\int_{-\infty}^\infty w_1\Big(\frac{x_1+i-u_1}{h}\Big)\frac{\partial}{\partial u_1}g(u_1,u_2)du_1\Bigg)du_2\\
    &= -\frac{1}{h^3} \sum_{i=1}^\infty \sum_{j=1}^\infty \int_{-\infty}^\infty w_1\Big(\frac{x_1+i-u_1}{h}\Big) \Bigg(\int_{-\infty}^\infty w'_2\Big(\frac{x_2+j-u_2}{h}\Big)\frac{\partial}{\partial u_1}g(u_1,u_2)du_2\Bigg)du_1\\
    &= \frac{1}{h^2} \sum_{i=1}^\infty \sum_{j=1}^\infty \int_{-\infty}^\infty \int_{-\infty}^\infty w_1\Big(\frac{x_1+i-u_1}{h}\Big) w_2\Big(\frac{x_2+j-u_2}{h}\Big)\frac{\partial^2}{\partial u_1\partial u_2}g(u_1,u_2)du_1du_2.
\end{align*}
By the substitutions $v_1:=u_1-i$ and $v_2:=u_2-j$ we get
\begin{align}\label{exp4}
    &\ex f_{nh}^{++}(x_1,x_2)\nonumber\\
    &=\frac{1}{h^2} \sum_{i=1}^\infty \sum_{j=1}^\infty \int_{-\infty}^\infty \int_{-\infty}^\infty w_1\Big(\frac{x_1-v_1}{h}\Big) w_2\Big(\frac{x_2-v_2}{h}\Big)\frac{\partial^2}{\partial u_1\partial u_2}g(v_1+i,v_2+j)dv_1dv_2.
\end{align}
Now we need to interchange integrals and sums again. Therefore, rewrite the equation above as
\begin{align}
    &\sum_{i=1}^\infty \sum_{j=1}^\infty \int_{-\infty}^\infty \int_{-\infty}^\infty w_1\Big(\frac{x_1-v_1}{h}\Big) w_2\Big(\frac{x_2-v_2}{h}\Big) \frac{\partial^2}{\partial v_1\partial v_2}g(v_1+i,v_2+j)dv_1dv_2\nonumber\\
    &=\lim_{M_1\to\infty}\lim_{M_2\to\infty}\sum_{i=1}^{M_1} \sum_{j=1}^{M_2} \int_{-\infty}^\infty \int_{-\infty}^\infty w_1\Big(\frac{x_1-v_1}{h}\Big) w_2\Big(\frac{x_2-v_2}{h}\Big) \frac{\partial^2}{\partial v_1\partial v_2}g(v_1+i,v_2+j)dv_1dv_2\nonumber\\
    &=\lim_{M_1\to\infty}\lim_{M_2\to\infty} \int_{-\infty}^\infty \int_{-\infty}^\infty w_1\Big(\frac{x_1-v_1}{h}\Big) w_2\Big(\frac{x_2-v_2}{h}\Big) \sum_{i=1}^{M_1} \sum_{j=1}^{M_1}\frac{\partial^2}{\partial v_1\partial v_2}g(v_1+i,v_2+j)dv_1dv_2.
\end{align}
By (\ref{invplus}) we have
$g(v_1+i,v_2)=F(v_1+i,v_2)-F(v_1+i,v_2-1)-F(v_1+i-1,v_2)+F(v_1+i-1,v_2-1)$,
so
\begin{equation*}
    \frac{\partial^2}{\partial v_1\partial v_2}g(v_1+i,v_2)=f(v_1+i,v_2)-f(v_1+i,v_2-1)-f(v_1+i-1,v_2)+f(v_1+i-1,v_2-1).
\end{equation*}
Following the summation of (\ref{invplus2}), we find
\begin{align}
    &\sum_{i=1}^{M_1}\frac{\partial^2}{\partial v_1\partial v_2}g(v_1+i,v_2)\nonumber\\
    &=\sum_{i=1}^{M_1}\{f(v_1+i,v_2)-f(v_1+i,v_2-1)-f(v_1+i-1,v_2)+f(v_1+i-1,v_2-1)\}\nonumber\\
    &=\sum_{i=1}^{M_1}\{f(v_1+i,v_2)-f(v_1+i-1,v_2)\}+\sum_{i=1}^{M_1}\{f(v_1+i-1,v_2-1)-f(v_1+i,v_2-1)\}\nonumber\\
    &=f(v_1+M_1,v_2)-f(v_1,v_2)-f(v_1,v_2-1)-f(v_1+M_1,v_2-1)
\end{align}
and
\begin{align}
    &\sum_{j=1}^{M_2}\sum_{i=1}^{M_1}\frac{\partial^2}{\partial v_1\partial v_2}g(v_1+i,v_2+j)\nonumber\\
    &=\sum_{j=1}^{M_2}\{f(v_1+M_1,v_2+j)-f(v_1,v_2+j)-f(v_1,v_2+j-1)-f(v_1+M_1,v_2+j-1)\}\nonumber\\
    &=\sum_{j=1}^{M_2}\{f(v_1+M_1,v_2+j)-f(v_1+M_1,v_2+j-1)+\sum_{j=1}^{M_2}\{f(v_1,v_2+j-1)-f(v_1,v_2+j)\}\nonumber\\
    &=f(v_1+M_1,v_2+M_2)-f(v_1+M_1,v_2)+f(v_1,v_2)-f(v_1,v_2+M_2).
\end{align}
Note that this sum is finite for all $v_1,v_2$, because $f$ is
bounded. Also note that changing the order of summation is allowed,
because $M_1,M_2<\infty$. By Lemma \ref{Inversion} we have
\begin{equation}
    \lim_{M_1\to\infty}\lim_{M_2\to\infty}\sum_{i=1}^{M_1}\sum_{j=1}^{M_2}\frac{\partial^2}{\partial v_1\partial v_2}g(v_1+i,v_2+j)=f(v_1,v_2)<\infty.
\end{equation}
We have assumed that $f$ is bounded, so let $f(v_1,v_2)\leq \frac{1}{4}A$ for all $v_1,v_2$, where $A>0$ is a constant. Observe the following inequality
\begin{align}
    &|f(v_1+M_1,v_2+M_2)-f(v_1+M_1,v_2)+f(v_1,v_2)-f(v_1,v_2+M_2)|\nonumber\\
    &\leq |f(v_1+M_1,v_2+M_2)|+|f(v_1+M_1,v_2)|+|f(v_1,v_2)|+|f(v_1,v_2+M_2)|\nonumber\\
    &\leq A,
\end{align}
for all $v_1,v_2,M_1$, and $M_2$. Note that, because $w_1$ and $w_2$
are   nonnegative, bounded and have bounded support,
\begin{equation}
    \int_{-\infty}^\infty \int_{-\infty}^\infty w_1\Big(\frac{x_1-v_1}{h}\Big) w_2\Big(\frac{x_2-v_2}{h}\Big)dv_1dv_2<\infty
\end{equation}
for all $x_1,x_2$. Thus we can apply the Lebesgue Dominated Convergenge Theorem to (\ref{exp4}), and find
\begin{align}
    &\lim_{M_1\to\infty}\lim_{M_2\to\infty} \int_{-\infty}^\infty \int_{-\infty}^\infty w_1\Big(\frac{x_1-v_1}{h}\Big) w_2\Big(\frac{x_2-v_2}{h}\Big) \sum_{i=1}^{M_1} \sum_{j=1}^{M_2}\frac{\partial^2}{\partial v_1\partial v_2}g(v_1+i,v_2+j)dv_1dv_2\nonumber\\
    &=\int_{-\infty}^\infty \int_{-\infty}^\infty w_1\Big(\frac{x_1-v_1}{h}\Big) w_2\Big(\frac{x_2-v_2}{h}\Big) \lim_{M_1\to\infty}\lim_{M_2\to\infty} \sum_{i=1}^{M_1} \sum_{j=1}^{M_2}\frac{\partial^2}{\partial v_1\partial v_2}g(v_1+i,v_2+j)dv_1dv_2\nonumber\\
    &=\int_{-\infty}^\infty \int_{-\infty}^\infty w_1\Big(\frac{x_1-v_1}{h}\Big) w_2\Big(\frac{x_2-v_2}{h}\Big) f(v_1,v_2)dv_1dv_2.
\end{align}
Summarizing  we now have
\begin{equation}
    \ex f_{nh}^{++}(x_1,x_2)=\frac{1}{h^2} \int_{-\infty}^\infty \int_{-\infty}^\infty w_1\Big(\frac{x_1-v_1}{h}\Big) w_2\Big(\frac{x_2-v_2}{h}\Big)f(v_1,v_2)dv_1dv_2.
\end{equation}
Substituting $z_1:=\frac{x_1-v_1}{h}$ and $z_2:=\frac{x_2-v_2}{h}$  we get
\begin{equation}
    \ex f_{nh}^{++}(x_1,x_2)=\int_{-\infty}^\infty \int_{-\infty}^\infty w_1(z_1)w_2(z_2)f(x_1-hz_1,x_2-hz_2)dz_1dz_2.
\end{equation}
Using the multivariate version of Taylor's theorem derived in Wand and Jones (1995) for this particular application, allows us to rewrite
\begin{align*}
     f(x_1-hz_1,&x_2-hz_2)  =f(x_1,x_2)-h(z_1f_{1}+z_2f_{2})(x_1,x_2)\\&+\frac{1}{2}h^2(z_1^2f_{11}+z_1z_2(f_{12}+f_{21})+z_2^2f_{22})(x_1,x_2)
    +o(h^2).
\end{align*}
We now obtain
\begin{align*}
    \ex f_{nh}^{++}(x_1,x_2)=&\int_{-\infty}^\infty \int_{-\infty}^\infty w_1(z_1)w_2(z_2)\Big(f(x_1,x_2)-h(z_1f_{1}+z_2f_{2})(x_1,x_2)\\
    &+\frac{1}{2}h^2(z_1^2f_{11}+z_1z_2(f_{12}+f_{21})+z_2^2f_{22})(x_1,x_2)+o(h^2)\Big)dz_1dz_2\\
    =&f(x_1,x_2)-hf_{1}(x_1,x_2)\int_{-\infty}^\infty \int_{-\infty}^\infty z_1w_1(z_1)w_2(z_2)dz_1dz_2\\
    &-hf_{2}(x_1,x_2)\int_{-\infty}^\infty \int_{-\infty}^\infty z_2w_1(z_1)w_2(z_2)dz_1dz_2\\
    &+\frac{1}{2}h^2f_{11}(x_1,x_2)\int_{-\infty}^\infty \int_{-\infty}^\infty z_1^2w_1(z_1)w_2(z_2)dz_1dz_2\\
    &+\frac{1}{2}h^2(f_{12}+f_{21})(x_1,x_2)\int_{-\infty}^\infty \int_{-\infty}^\infty z_1z_2w_1(z_1)w_2(z_2)dz_1dz_2\\
    &+\frac{1}{2}h^2f_{22}(x_1,x_2)\int_{-\infty}^\infty \int_{-\infty}^\infty z_2^2w_1(z_1)w_2(z_2)dz_1dz_2+o(h^2)\\
    =&f(x_1,x_2)+\frac{1}{2}h^2\Big(\int_{-\infty}^\infty z^2w_1(z)dz f_{11}(x_1,x_2)+\int_{-\infty}^\infty z^2w_2(z)dzf_{22}(x_1,x_2)\Big)+o(h^2).
\end{align*}
This proves statement (\ref{expectation}) of the theorem for this individual estimator.

It is easily seen that
\begin{align*}
    &\ex f^{--}_{nh}(x_1,x_2)=\ex f^{-+}_{nh}(x_1,x_2)=\ex f^{+-}_{nh}(x_1,x_2)=\ex f^{++}_{nh}(x_1,x_2)\\
    &=f(x_1,x_2)+\frac{1}{2}h^2\Big(\int_{-\infty}^\infty z^2w_1(z)dz f_{11}(x_1,x_2)+\int_{-\infty}^\infty z^2w_2(z)dzf_{22}(x_1,x_2)\Big)+o(h^2)
\end{align*}
and thus $$\ex
f_{nh}^{(t)}(x_1,x_2)=f(x_1,x_2)+\frac{1}{2}h^2\Big(\int_{-\infty}^\infty
z^2w_1(z)dz f_{11}(x_1,x_2)+\int_{-\infty}^\infty
z^2w_2(z)dzf_{22}(x_1,x_2)\Big)+o(h^2),$$ proving equation
(\ref{expectation}).

Next let us derive the asymptotic variance. First, define
\begin{equation}
    U_{kh}^{++}(x_1,x_2):= \frac{1}{h^4}\sum_{i=1}^\infty \sum_{j=1}^\infty w'_1\bigg(\frac{x_1+i-X_{k1}}{h}\bigg) w'_2\bigg(\frac{x_2+j-X_{k2}}{h}\bigg).
\end{equation}
Then $f_{nh}^{++}(x_1,x_2) = \frac{1}{n}\sum_{k=1}^n U_{kh}^{++}(x_1,x_2)$, and since the terms $U_{kh}^{++}$ are independent,
\begin{equation}
    \Var(f_{nh}^{++}(x_1,x_2))= \frac{1}{n}\Var(U_{1h}^{++}(x_1,x_2)).
\end{equation}
Secondly, we will determine the variance of $U_{1h}^{++}(x_1,x_2)$. We have
\begin{equation}
    \Var(U_{1h}^{++}(x_1,x_2))=\ex U_{1h}^{++}(x_1,x_2)^2 - (\ex U_{1h}^{++}(x_1,x_2))^2.
\end{equation}
Let us begin with determining $\ex U_{1h}^{++}(x_1,x_2)^2$. Note that, if $h<\frac{1}{2}$, we have
\begin{equation}\label{product}
    w'_1\bigg(\frac{x_1+i_1-X_{k1}}{h}\bigg)w'_2\bigg(\frac{x_2+i_2-X_{k2}}{h}\bigg)w'_1\bigg(\frac{x_1+j_1-X_{k1}}{h}\bigg)w'_2\bigg(\frac{x_2+j_2-X_{k2}}{h}\bigg)=0
\end{equation}
unless $i_1=i_2$ and $j_1=j_2$, where $i_1,i_2,j_1,j_2\in \mathbb Z$. This holds because if $i_1\neq i_2$ or $j_1\neq j_2$, then at least two pairs of arguments in the product (\ref{product}) are more than distance two apart, rendering the product equal to zero. Thus in the following equation, as $h \to 0$, only the square products do not vanish and we can write
\begin{align*}
    \ex U_{1h}^{++}(x_1,x_2)^2&=\ex \bigg(\frac{1}{h^4} \sum_{i=1}^\infty \sum_{j=1}^\infty w'_1\bigg(\frac{x_1+i-X_{11}}{h}\bigg) w'_2\bigg(\frac{x_2+j-X_{12}}{h}\bigg)\bigg)^2\\
    &=\frac{1}{h^8} \sum_{i=1}^\infty \sum_{j=1}^\infty \ex \bigg(w'_1\bigg(\frac{x_1+i-X_{11}}{h}\bigg) w'_2\bigg(\frac{x_2+j-X_{12}}{h}\bigg)\bigg)^2.
\end{align*}
Now we use the substitutions $v_1:=u_1-i$ and $v_2:=u_2-j$ to obtain
\begin{align*}
    \ex U_{1h}^{++}(x_1,x_2)^2&=\frac{1}{h^8} \sum_{i=1}^\infty \sum_{j=1}^\infty \int_{-\infty}^\infty \int_{-\infty}^\infty \Big(w'_1\Big(\frac{x_1+i-u_1}{h}\Big) w'_2\Big(\frac{x_2+j-u_2}{h}\Big)\Big)^2 g(u_1,u_2)du_1du_2\\
    &=\frac{1}{h^8} \sum_{i=1}^\infty \sum_{j=1}^\infty \int_{-\infty}^\infty \int_{-\infty}^\infty \Big(w'_1\bigg(\frac{x_1-v_1}{h}\Big) w'_2\Big(\frac{x_2-v_2}{h}\Big)\Big)^2 g(v_1+i,v_2+j)dv_1dv_2.
\end{align*}
Note that the integrand is nonnegative, thus interchanging sums and integrals is allowed (Fubini), so
\begin{align*}
    \ex U_{1h}^{++}(x_1,x_2)^2&=\frac{1}{h^8} \int_{-\infty}^\infty \int_{-\infty}^\infty\Big(w'_1\Big(\frac{x_1-v_1}{h}\Big) w'_2\Big(\frac{x_2-v_2}{h}\Big)\Big)^2 \sum_{i=1}^\infty \sum_{j=1}^\infty g(v_1+i,v_2+j)dv_1dv_2\\
    &=\frac{1}{h^8} \int_{-\infty}^\infty \int_{-\infty}^\infty w'_1\Big(\frac{x_1-v_1}{h}\Big)^2 w'_2\Big(\frac{x_2-v_2}{h}\Big)^2 F^{++}(v_1,v_2)dv_1dv_2.
\end{align*}
Now apply the substitutions $z_1=(x_1-v_1)/h$ and $z_2=(x_2-v_2)/h$
and recall the bounded support of $w'_1$ and $w'_2$. Furthermore,
because $\lim_{h\to 0}
F^{++}(x_1-hz_1,x_2-hz_2)=F^{++}(x_1,x_2)\le1$, we can again apply
the Lebesgue dominated convergence theorem
\begin{align}
    \ex U_{1h}^{++}(x_1,x_2)^2&=\frac{1}{h^6} \int_{-1}^1 \int_{-1}^1 w'_1(z_1)^2 w'_2(z_2)^2 F^{++}(x_1-hz_1,x_2-hz_2)dz_1dz_2\nonumber
 \\
    &=\frac{1}{h^6} F^{++}(x_1,x_2)\int_{-1}^1
    w'_1(z_1 )^2dz_1
\int_{-1}^1w'_2(z_2 )^2 dz_2 +o(h^{-6}).\label{88}
\end{align}
Now note that $\ex U^{++}_{1h}(x_1,x_2)=\ex f^{++}_{nh}(x_1,x_2)=f(x_1,x_2)+O(h^2)$. So
\begin{align*}
    \Var(f_{nh}^{++}(x_1,x_2))&=\frac{1}{n}\Var(U_{1h}^{++}(x_1,x_2))\\
    &=\frac{1}{n}\Big[\ex U_{1h}^{++}(x_1,x_2)^2 - (\ex U_{1h}^{++}(x_1,x_2))^2\Big]\\
    &=\frac{1}{n}\Bigg[\frac{1}{h^6}F^{++}(x_1,x_2)\int_{-1}^1
    w'_1(z_1)^2dz_1
\int_{-1}^1w'_2(z_2)^2 dz_2+o(h^{-6})-f(x_1,x_2)^2-O(h^2)\Bigg]\\
    &=\frac{1}{nh^6}F^{++}(x_1,x_2)\int_{-1}^1
    w'_1(z_1)^2dz_1
\int_{-1}^1w'_2(z_2)^2 dz_2+o(n^{-1}h^{-6}).
\end{align*}
We can follow a similar procedure to obtain the variances of the other estimators. To summarize we get
\begin{align*}
&\Var(f_{nh}^{--}(x_1,x_2)) =
\frac{1}{nh^6}F^{--}(x_1,x_2)\int_{-1}^1
    w'_1(z_1)^2dz_1
\int_{-1}^1w'_2(z_2)^2 dz_2+o(n^{-1}h^{-6}),\\
&\Var(f_{nh}^{-+}(x_1,x_2)) =
\frac{1}{nh^6}F^{-+}(x_1,x_2)\int_{-1}^1
    w'_1(z_1)^2dz_1
\int_{-1}^1w'_2(z_2)^2 dz_2+o(n^{-1}h^{-6}),\\
&\Var(f_{nh}^{+-}(x_1,x_2)) =
\frac{1}{nh^6}F^{+-}(x_1,x_2)\int_{-1}^1
    w'_1(z_1)^2dz_1
\int_{-1}^1w'_2(z_2)^2 dz_2+o(n^{-1}h^{-6}),\\
&\Var(f_{nh}^{++}(x_1,x_2)) =
\frac{1}{nh^6}F^{++}(x_1,x_2)\int_{-1}^1
    w'_1(z_1)^2dz_1
\int_{-1}^1w'_2(z_2)^2 dz_2+o(n^{-1}h^{-6}).
\end{align*}

Now let us determine the variance of combinations of these estimators. We have
\begin{align*}
    \Var&(f_{nh}^{(t)}(x_1,x_2))=
    \Var(t_1f^{--}_{nh}(x_1,x_2)+t_2f^{-+}_{nh}(x_1,x_2)+t_3f^{+-}_{nh}(x_1,x_2)+t_4f^{++}_{nh}(x_1,x_2)) \\
    =&t_1^2\Var(f^{--}_{nh}(x_1,x_2))+t_2^2\Var(f^{-+}_{nh}(x_1,x_2))+t_3^2\Var(f^{+-}_{nh}(x_1,x_2))+t_4^2\Var(f^{++}_{nh}(x_1,x_2))\nonumber\\
    &+2t_1t_2\Cov(f^{--}_{nh}(x_1,x_2),f^{-+}_{nh}(x_1,x_2))+2t_1t_3\Cov(f^{--}_{nh}(x_1,x_2),f^{+-}_{nh}(x_1,x_2)) \\
    &+2t_1t_4\Cov(f^{--}_{nh}(x_1,x_2),f^{++}_{nh}(x_1,x_2))+2t_2t_3\Cov(f^{-+}_{nh}(x_1,x_2),f^{+-}_{nh}(x_1,x_2)) \\
    &+2t_2t_4\Cov(f^{-+}_{nh}(x_1,x_2),f^{++}_{nh}(x_1,x_2))+2t_3t_4\Cov(f^{+-}_{nh}(x_1,x_2),f^{++}_{nh}(x_1,x_2)).
\end{align*}
Let us look at $\Cov(f^{--}_{nh}(x_1,x_2),f^{-+}_{nh}(x_1,x_2))$. In similar fashion as we determined the variance, we find
\begin{align*}
    \Cov(f^{--}_{nh}(x_1,x_2),f^{-+}_{nh}(x_1,x_2))&=\frac{1}{n}\Cov(U_{1h}^{--}(x_1,x_2),U_{1h}^{-+}(x_1,x_2))\\
    &=\frac{1}{n}\big[\ex U_{1h}^{--}(x_1,x_2)U_{1h}^{-+}(x_1,x_2) - \ex U_{1h}^{--}(x_1,x_2)\ex U_{1h}^{-+}(x_1,x_2)\big]
\end{align*}
Let us first determine $\ex U_{1h}^{--}(x_1,x_2)U_{1h}^{-+}(x_1,x_2)$. Note that, if $h<\frac{1}{2}$, we have
\begin{equation}\label{product2}
    w'_1\bigg(\frac{x_1-i_1-X_{k1}}{h}\bigg)w'_2\bigg(\frac{x_2-i_2-X_{k2}}{h}\bigg)w'_1\bigg(\frac{x_1-j_1-X_{k1}}{h}\bigg)w'_2\bigg(\frac{x_2+j_2-X_{k2}}{h}\bigg)=0,
\end{equation}
for all $i_1,i_2,j_1$ and $j_2$. This holds because the second and fourth argument in the product (\ref{product2}) are always more than distance two apart, rendering the product equal to zero. Thus
\begin{equation}
    \ex U_{1h}^{--}(x_1,x_2)U_{1h}^{-+}(x_1,x_2)=0.
\end{equation}
Secondly, because we have already determined $\ex U_{1h}^{--}(x_1,x_2)$ and $\ex U_{1h}^{-+}(x_1,x_2)$ earlier, we know that
\begin{equation}
    \ex U_{1h}^{--}(x_1,x_2)\ex U_{1h}^{-+}(x_1,x_2)=f(x_1,x_2)^2+O(h^2).
\end{equation}
Thus
\begin{equation}
    \Cov(f^{--}_{nh}(x_1,x_2),f^{-+}_{nh}(x_1,x_2))=\frac{1}{n}[-f(x_1,x_2)^2-O(h^2)]=o(n^{-1}h^{-6}).
\end{equation}
This result holds for all the covariances. So we arrive at
\begin{align*}
    \Var(f_{nh}(x_1,x_2))=&(t_1^2F^{--}(x_1,x_2)+t_2^2F^{-+}(x_1,x_2)+t_3^2F^{+-}(x_1,x_2)+t_4^2F^{++}(x_1,x_2))\\
    &\frac{1}{nh^6}\int_{-1}^1
    w'_1(z_1)^2dz_1
\int_{-1}^1w'_2(z_2)^2 dz_2+o(n^{-1}h^{-6})\\
    =&B(x_1,x_2,t_1,t_2,t_3,t_4)\frac{1}{nh^6}\int_{-1}^1
    w'_1(z_1)^2dz_1
\int_{-1}^1w'_2(z_2)^2 dz_2+o(n^{-1}h^{-6}).
\end{align*}
This proves statement (\ref{variance1}) of the theorem.\hfill$\Box$

\subsection{ Proof of Theorem \ref{mainthmest}}

The convex combination of the four  density estimators  is given by
\begin{equation}\label{convest}
    f_{nh}^{(t )}(x_1,x_2)=t_1 f^{--}_{nh}(x_1,x_2)+t_2 f^{-+}_{nh}(x_1,x_2)+t_3 f^{+-}_{nh}(x_1,x_2)+t_4 f^{++}_{nh}(x_1,x_2),
\end{equation}
where $t_1+t_2+t_3+t_4=1$.
Now define
\begin{align*}
&S_{1nh} (x_1,x_2)= f^{--}_{nh}(x_1,x_2) -f^{+-}_{nh}(x_1,x_2) ,\\
&S_{2nh} (x_1,x_2)=  -f^{-+}_{nh}(x_1,x_2) + f^{++}_{nh}(x_1,x_2),\\
&S_{3nh} (x_1,x_2)= f^{--}_{nh}(x_1,x_2)-f^{-+}_{nh}(x_1,x_2) ,\\
&S_{4nh} (x_1,x_2)=  -f^{+-}_{nh}(x_1,x_2)+ f^{++}_{nh}(x_1,x_2).\\
\end{align*}
We can rewrite (\ref{convest}) as
\begin{equation}\label{convest2}
    f_{nh}^{(t)}(x_1,x_2)=f^{--}_{nh}(x_1,x_2)-(t_3+t_4)S_{1nh} (x_1,x_2)-t_2S_{3nh} (x_1,x_2)+ t_4S_{4nh}(x_1,x_2),
\end{equation}
\begin{lem}\label{terms}
Under the conditions of Theorem \ref{mainthmest} we have, for $i=1,\ldots,4$,
\begin{align}
&\ex S_{inh} (x_1,x_2)=0,\\
&\ex S_{inh} (x_1,x_2)^2=O\Big(\frac{1}{nh^6}\Big),\\
&\ex S_{inh} (x_1,x_2)^4=O\Big(\frac{1}{n^2h^{12}}\Big).
\end{align}
\end{lem}

\noindent{\bf Proof}

We give the proof for $S_{1nh} (x_1,x_2)$. The other claims can be proved similarly.

Note that
\begin{align*}
S_{1nh} (x_1,x_2)&= f^{--}_{nh}(x_1,x_2) -f^{+-}_{nh}(x_1,x_2)\\
&=\frac{1}{nh^4} \sum_{k=1}^n \sum_{i=0}^\infty \sum_{j=0}^\infty w'_1(\frac{x_1-i-X_{k1}}{h}) w'_2(\frac{x_2-j-X_{k2}}{h})\\
&\quad\quad+\frac{1}{nh^4} \sum_{k=1}^n \sum_{i=1}^\infty \sum_{j=0}^\infty w'_1(\frac{x_1+i-X_{k1}}{h}) w'_2(\frac{x_2-j-X_{k2}}{h})\\
&=\frac{1}{nh^4} \sum_{k=1}^n \sum_{i=-\infty}^\infty
\sum_{j=0}^\infty w'_1(\frac{x_1-i-X_{k1}}{h})
w'_2(\frac{x_2-j-X_{k2}}{h}).
\end{align*}
Define
\begin{equation}
    U_{1kh} (x_1,x_2):= \frac{1}{h^4}\sum_{i=-\infty}^\infty \sum_{j=1}^\infty w'_1\bigg(\frac{x_1+i-X_{k1}}{h}\bigg) w'_2\bigg(\frac{x_2+j-X_{k2}}{h}\bigg).
\end{equation}
Then $S_{1nh} (x_1,x_2) = \frac{1}{n}\sum_{k=1}^n U_{1kh} (x_1,x_2)$ and the terms in the sum are independent.

Following similar steps as in the proof of Theorem \ref{mainthm1} we get
\begin{align*}
\ex U_{1kh} &(x_1,x_2)=\int_{-\infty}^\infty \int_{-\infty}^\infty w_1\Big(\frac{x_1-v_1}{h}\Big) w_2\Big(\frac{x_2-v_2}{h}\Big)  \frac{\partial^2}{\partial v_1\partial v_2}\sum_{i=-\infty}^{\infty} \sum_{j=1}^{\infty} g(v_1+i,v_2+j)dv_1dv_2\\
&=\int_{-\infty}^\infty \int_{-\infty}^\infty
w_1\Big(\frac{x_1-v_1}{h}\Big) w_2\Big(\frac{x_2-v_2}{h}\Big)
\frac{\partial^2}{\partial v_1\partial
v_2}(1-F_{Y_2}(v_2))dv_1dv_2=0.
\end{align*}
We also have, as in the same proof,
$$
\ex S_{1nh} (x_1,x_2)^2=\var(S_{1nh} (x_1,x_2))=\frac{1}{n}\, \var(U_{11h} (x_1,x_2))=O\Big(\frac{1}{nh^6}\Big).
$$

Finally we consider the fourth moment of $S_{1nh} (x_1,x_2) = \frac{1}{n}\sum_{k=1}^n U_{1kh} (x_1,x_2)$. By independence of the terms we have
\begin{align*}
\ex S_{1nh}& (x_1,x_2)^4=\frac{1}{n^3}\, \ex U_{11h} (x_1,x_2)^4 + \frac{3(n-1)}{n^3}\, \Big(\ex U_{11h} (x_1,x_2)^2\Big)^2\\
&=\frac{1}{n^3}\, O\Big(\frac{1}{h^{14}}\Big) + \frac{3(n-1)}{n^3}\,\Big(O\Big(\frac{1}{h^6}\Big)\Big)^2= O\Big(\frac{1}{n^2h^{12}}\Big) .
\end{align*}
This completes the proof of the lemma.
\hfill $\Box$
\bigskip

From (\ref{convest2}) we get, omitting the arguments $(x_1,x_2)$,
\begin{equation} \label{representation}
    f_{nh}^{(\hat t_n)}  -f_{nh}^{(\bar t)}
 = -(\hat t_{n3} - \bar t_3)S_{1nh}  - (\hat t_{n4} - \bar t_4)S_{1nh}    -(\hat t_{n2} -\bar t_2)S_{3nh}  + (\hat t_{n4}-\bar t_4)S_{4nh} .
\end{equation}
Hence, under the assumptions of the theorem and by the Cauchy Schwarz inequality, we have

\begin{align*}
\ex|f_{nh}^{(\hat t_n)}&  -f_{nh}^{(\bar t)}|
 \leq
\ex |\hat t_{n3} - \bar t_3||S_{1nh}|  + \ex |  \hat t_{n4} - \bar
t_4||S_{1nh}|
+\ex |\hat t_{n2} -\bar t_2||S_{3nh}|  + \ex |\hat t_{n4}-\bar t_4||S_{4nh}|\\
&\leq \Big(\ex (\hat t_{n3} - \bar t_3) ^2\Big)^{1/2}\Big(\ex S_{1nh}^2\Big)^{1/2}  +
\Big(\ex (\hat t_{n4} - \bar t_4) ^2\Big)^{1/2}\Big(\ex S_{1nh}^2\Big)^{1/2}     \\
&+\Big(\ex (\hat t_{n2} - \bar t_2) ^2\Big)^{1/2}\Big(\ex S_{3nh}^2\Big)^{1/2}  +
\Big(\ex (\hat t_{n4} - \bar t_4) ^2\Big)^{1/2}\Big(\ex S_{4nh}^2\Big)^{1/2}\\
&=\Big(o(n h^{10}) O\Big(\frac{1}{nh^6}\Big)\Big)^{1/2}= o(h^2).
\end{align*}
Similarly we have, since $(y_1+y_2+y_3+y_4)^2\leq 4(y_1^2+y_2^2+y_3^2+y_4^2)$,
\begin{align*}
\var(f_{nh}^{(\hat t_n)}&  -f_{nh}^{(\bar t)})\leq \ex(f_{nh}^{(\hat t_n)}   -f_{nh}^{(\bar t)})^2 \\
& \leq
4\ex (\hat t_{n3} - \bar t_3)^2 S_{1nh}^2  + 4\ex (\hat t_{n4} - \bar t_4)^2S_{1nh}^2
+4\ex (\hat t_{n2} -\bar t_2)^2S_{3nh}^2  + 4\ex (\hat t_{n4}-\bar t_4)^2S_{4nh}^2\\
& \leq 4\Big(\ex (\hat t_{n3} - \bar t_3) ^4\Big)^{1/2}\Big(\ex S_{1nh}^4\Big)^{1/2}  +
4\Big(\ex (\hat t_{n4} - \bar t_4) ^4\Big)^{1/2}\Big(\ex S_{1nh}^4\Big)^{1/2}     \\
&+4\Big(\ex (\hat t_{n2} - \bar t_2) ^4\Big)^{1/2}\Big(\ex S_{3nh}^4\Big)^{1/2}  +4
\Big(\ex (\hat t_{n4} -\bar  t_4) ^4\Big)^{1/2}\Big(\ex S_{4nh}^4\Big)^{1/2}\\
&=o(1) \Big(O\Big(\frac{1}{n^2h^{12}}\Big)\Big)^{1/2}= o\Big(\frac{1}{nh^6}\Big).
\end{align*}
Since the two bounds above are negligible compared to the order of the bias and variance in Theorem \ref{mainthm1}
it follows that this theorem also holds for the estimator with estimated weights.

In order to prove asymptotic normality note that by Lemma \ref{terms}  and condition (\ref{c3}) it follows
that $\sqrt{n}h^3$  times each of the terms in the representation (\ref{representation}) vanish in probability.
Also it follows that $\sqrt{n}h^3$ times  the expectation of (\ref{representation}) vanishes asymptotically. Hence the limit distributions of of $\sqrt{n}h^3(f_{nh}^{(\hat t_n)}- \ex f_{nh}^{(\hat t_n)})$ and $\sqrt{n}h^3( f_{nh}^{(\bar t)} - \ex f_{nh}^{(\bar t)})$
coincide. The limit distribution of the latter follows by checking the Lyapounov condition for asymptotic normality.

\hfill $\Box$

\subsection{Proof of lemma \ref{checkest}}

\noindent{\bf Proof}
Let us first introduce some notation.
Define the vectors ${\mathbf v}(x_1,x_2)$ and ${\mathbf {\tilde v}}_{n\tilde h}(x_1,x_2) $ by
\begin{align*}
& {\mathbf v}(x_1,x_2)=( F^{--}(x_1,x_2), F^{-+}(x_1,x_2), F^{+-}(x_1,x_2),
F^{++}(x_1,x_2)),\\
& {\mathbf {\tilde v}}_{n\tilde h}(x_1,x_2)=({\tilde F}_{n\tilde h}^{--}(x_1,x_2), {\tilde F}_{n\tilde h}^{-+}(x_1,x_2), {\tilde F}_{n\tilde h}^{+-}(x_1,x_2),
{\tilde F}_{n\tilde h}^{++}(x_1,x_2)) .
\end{align*}
Note that, for $n$ large enough,  the components of these vectors are all at least   $\epsilon_n$ and that they are at most one.

We will only check (\ref{c1}) and (\ref{c2}) for $i$ equal to one.
The other cases can be treated similarly.
Then we also need the vector of partial derivatives of the the function  $\tilde t_1(y_1,y_2, ,y_3 ,y_4)$.
Note that on the line segment between ${\mathbf {\tilde v}}_{n\tilde h}(x_1,x_2) $ and ${\mathbf v}(x_1,x_2)$ all the components
are all at least   $\epsilon_n$ and that they are at most one. This implies after some computation
$$
\| \nabla \tilde t_1(y_1,y_2, ,y_3 ,y_4)\| ^2\leq   \frac{B}{ \epsilon_n^6},
$$
for some constant $B$, for all points $(y_1,y_2, ,y_3 ,y_4)$ on this line segment.

We can now apply the multivariate mean value theorem and the Cauchy Schwarz inequality to get
\begin{align*}
(\hat t_{n1}(x_1,x_2) &-\bar t_1(x_1,x_2))^2=(\tilde t_1({\mathbf {\tilde v}}_{n\tilde h}(x_1,x_2))
-\tilde t_1({\mathbf v}(x_1,x_2)))^2\\
&=({\mathbf {\tilde v}}_{n\tilde h}(x_1,x_2)-{\mathbf v}(x_1,x_2))\cdot
\nabla \tilde t_1(y_1,y_2, ,y_3 ,y_4))^2\\
&\leq \|{\mathbf {\tilde v}}_{n\tilde h}(x_1,x_2)-{\mathbf v}(x_1,x_2) \|^2\| \nabla \tilde t_1(y_1,y_2, ,y_3 ,y_4))\|^2\\
&\leq \frac{B}{ \epsilon_n^6}\, \|{\mathbf {\tilde v}}_{n\tilde h}(x_1,x_2)-{\mathbf v}(x_1,x_2) \|^2,
\end{align*}
where $(y_1,y_2, ,y_3 ,y_4)$ is a point  on the line segment between
${\mathbf {\tilde v}}(x_1,x_2)_{n\tilde h}$ and ${\mathbf
v}(x_1,x_2)$. Note that  $\|{\mathbf {\tilde v}}_{n\tilde
h}(x_1,x_2)-{\mathbf v}(x_1,x_2) \|^2$ is a sum of four terms like
$({\tilde F}_{n\tilde h}^{--}(x_1,x_2)-F^{--}(x_1,x_2))^2$, which is
smaller than  $( F_{n\tilde h}^{--}(x_1,x_2)-F^{--}(x_1,x_2))^2$,
and that $\ex ( F_{n\tilde h}^{--}(x_1,x_2)-F^{--}(x_1,x_2))^2$
equals  the variance plus the squared bias of $F_{n\tilde
h}^{--}(x_1,x_2)$. By Theorem \ref{mainthm2} we can bound these to
get
$$
\ex (\hat t_{n1}(x_1,x_2)  -\bar t_1(x_1,x_2))^2\leq \frac{B}{ \epsilon_n^6}\Big( O\Big(\frac{1}{n\tilde h^2}\Big)+O(\tilde h^4)\Big)= O(n^{-2/3} (\log n)^6),
$$
for a bandwidth $h$ of order $n^{-1/6}$. This implies that (\ref{c1}) is satisfied.

Let us now check that (\ref{c2}) is satisfied.
By an argument similar to the one above it suffices to check if terms like
 $\ex(F_{n\tilde h}^{--}(x_1,x_2)-F^{--}(x_1,x_2))^4$ vanish asymptotically.
Write
$$
F_{n\tilde h}^{--}(x_1,x_2)-F^{--}(x_1,x_2)=F_{n\tilde h}^{--}(x_1,x_2)-\ex F_{n\tilde h}^{--}(x_1,x_2) +\ex F_{n\tilde h}^{--}(x_1,x_2)-F^{--}(x_1,x_2).
$$
By the triangle inequality we have
\begin{align*}
\Big(\ex(F_{n\tilde h}^{--}&(x_1,x_2) -F^{--}(x_1,x_2))^4\Big)^{1/4}\leq \\
&\leq  \Big(\ex(F_{n\tilde h}^{--}(x_1,x_2)-\ex F_{n\tilde
h}^{--}(x_1,x_2))^4\Big)^{1/4}+\Big(\ex F_{n\tilde
h}^{--}(x_1,x_2)-F^{--}(x_1,x_2)\Big) .
\end{align*}
So, by $(a+b)^4\leq 8(a^4+b^4), a,b\geq 0$,  we also have
\begin{align*}
 \ex(F_{n\tilde h}^{--}&(x_1,x_2)-F^{--}(x_1,x_2))^4 \leq \\
&\leq  8 \ex(F_{n\tilde h}^{--}(x_1,x_2)-\ex F_{n\tilde h}^{--}(x_1,x_2))^4  +8 \Big(\ex F_{n\tilde h}^{--}(x_1,x_2)-F^{--}(x_1,x_2)\Big)^4 .
\end{align*}
Since the bias vanishes by Theorem \ref{mainthm2} it suffices to prove the bound of the lemma for the fourth power of the error.

Recall from the proof of Theorem \ref{mainthm2} that
$F_{n\tilde h}^{--}(x_1,x_2) = \frac{1}{n}\sum_{k=1}^n U_{kh}^{--}(x_1,x_2)$, where
$$
    U_{k\tilde h}^{--}(x_1,x_2):= \frac{1}{\tilde h^2}\sum_{i=0}^\infty \sum_{j=0}^\infty w_1\bigg(\frac{x_1-i-X_{k1}}{\tilde h}\bigg) w_2\bigg(\frac{x_2-j-X_{k2}}{\tilde h}\bigg).
$$
Note that the $U_{k\tilde h}^{--}$ are independent.
Now write
$$
F_{n\tilde h}^{--}(x_1,x_2)-\ex F_{n\tilde h}^{--}(x_1,x_2)= \frac{1}{n}\sum_{k=1}^n \tilde U_{k\tilde h}^{--}(x_1,x_2),
$$
where $\tilde U_{k\tilde h}^{--}(x_1,x_2)=U_{k\tilde h}^{--}(x_1,x_2) -\ex U_{k\tilde h}^{--}(x_1,x_2)$.
 Since   $\ex \tilde U_{kh}^{--}(x_1,x_2)$ equals zero we have
$$
\ex\Big(\frac{1}{n}\sum_{i=1}^n \tilde U_{k\tilde h}^{--}(x_1,x_2)\Big)^4=\frac{1}{n^3}\ex\Big(\tilde U_{k\tilde h}^{--}(x_1,x_2)^4\Big) +
3\,\frac{n-1}{n^3} \Big(\ex\Big(\tilde U_{k\tilde h}^{--}(x_1,x_2)^2\Big)\Big)^2.
$$
Similar to the derivation of (\ref{88}) we get
$$
\frac{1}{n^3}\ex\Big(\tilde U_{k\tilde h}^{--}(x_1,x_2)^4\Big)\sim \frac{1}{n^3}\ex\Big( U_{k\tilde h}^{--}(x_1,x_2)^2\Big)\sim
O\Big(\frac{1}{n^3 \tilde h^4}\Big)$$
and
$$
3\,\frac{n-1}{n^3}\big(\ex\Big(\tilde U_{k\tilde h}^{--}(x_1,x_2)^2\Big)\Big)^2=
3\,\frac{n-1}{n^3}\big(\var (U_{k\tilde h}^{--}(x_1,x_2))\Big)^2\sim O\Big(\frac{1}{n^2\tilde h^{4}}\Big).
$$
Under the condition on $\tilde h$ in the lemma both terms vanish. This shows that (\ref{c2}) is satisfied as well.
Condition (\ref{c3}) follows from  condition (\ref{c2}) by the Cauchy-Schwarz inequality. \hfill $\Box$

\subsection{An inequality}

The next lemma can be used to derive the  weights that minimize the asymptotic variance of the convex combination of the original for estimators of the density $f$.

\begin{lem}\label{inequality}
Let $a_1,\ldots,a_m$ be $m$ positive numbers. Then for all positive $t_1,\ldots,t_m$ with
$t_1+\ldots + t_m =1$ we have
\begin{equation}\label{ineq}
a_1t_1^2+\ldots +a_mt_m^2\geq \frac{a_1a_2\dots a_m}{s_m(a_1,\ldots,a_m)},
\end{equation}
where $s_m(a_1,\ldots,a_m)$ is defined by
\begin{equation}
s_m(a_1,\ldots,a_m)=a_2a_3\dots a_m+\sum_{j=2}^{m-1} a_1\dots a_{j-1}a_{j+1}\dots a_m + a_1a_2\dots a_{m-1},
\end{equation}
the sum of the $m$ products of length $m-1$ obtained by skipping one term  in the full product.

The minimum is attained at the $t$ vector given by $t_1= a_2 a_3 \cdots  a_m/s_m(a_1,\ldots ,a_m)$ and $t_m= a_1 a_2 \cdots  a_{m-1}/s_m(a_1,\ldots ,a_m)$ and
\begin{equation*}
t_i=\frac{a_1 a_2 \cdots a_{i-1}a_{i+1}\cdots a_m}{s_m(a_1,\ldots ,a_m)}, \quad i=2,\ldots ,m-1.
\end{equation*}
\end{lem}

\medskip

\noindent {\bf Proof}
Introduce the inner product $<\cdot, \cdot >_a$ and corresponding norm $\| \cdot \|_a $ by
\begin{align}
<x, y >_a&= a_2a_3\dots a_m\,x_1y_1+a_1a_3\dots a_m\,x_2y_2+\ldots + a_1a_2\dots a_{m-1}\,x_my_m,\\
\| x \|_a  &= \Big(a_2a_3\dots a_m\,x_1^2+a_1a_3\dots a_m\,x_2^2+\ldots + a_1a_2\dots a_{m-1}\,x_m^2\Big)^{1/2}.
\end{align}
Then, with $\mathbf 1$ equal to the vector of $m$ ones, the Cauchy-Schwarz inequality implies
\begin{align*}
&a_1a_2\dots a_m= (a_1a_2\dots a_m)(t_1+t_2+\ldots +t_m)\\
&=< {\mathbf 1},(a_1t_1,a_2t_2,\ldots ,a_mt_m) >_a
\ \leq  \ \| 1 \|_a\| (a_1t_1,a_2t_2,\ldots ,a_mt_m) \|_a \\
&=\sqrt{s(a_1,\ldots,a_m)}\Big(a_2a_3\dots a_m\,(a_1t_1)^2+a_1a_3\dots a_m\,(a_2t_2)^2+\ldots + a_1a_2\dots a_{m-1}\,(a_mt_m)^2\Big)^{1/2}\\
&=\sqrt{s(a_1,\ldots,a_m)}\Big((a_1a_2\dots a_m)(a_1t_1^2+a_2t_2^2+ \ldots +  a_mt_m^2)\Big)^{1/2},
\end{align*}
which implies the inequality after some rewriting.
\hfill $\Box$

\end{document}